%% file: main.tex
\documentclass[journal,onecolumn]{IEEEtran}

\usepackage[utf8]{inputenc}
\usepackage{graphicx}
\usepackage{amsthm,amsmath,amsfonts,amssymb,mathrsfs, color}
\theoremstyle{definition}
\newtheorem{definition}{Definition}
\usepackage{subfigure}
\theoremstyle{remark}

\usepackage[normalem]{ulem}
\newtheoremstyle{mytheorem}{0.5cm}{0.2cm}{\slshape}{ }{\bfseries}{.}{ }{}
\theoremstyle{mytheorem}
\newtheorem{Th}{Theorem}
\newtheorem{Prop}[definition]{Proposition}
\newtheorem{Def}[definition]{Definition}

\newtheorem{Lem}[definition]{Lemma}
\newtheorem{Cor}[definition]{Corollary}
\newtheorem{Rem}[definition]{Remark}
\newtheorem{remark}[definition]{Remark}

\def\pgeq#1{\stackrel{\text{\scriptsize(#1)}}{\geq}}
\def\peq#1{\stackrel{\text{\scriptsize(#1)}}{=}}
\title{Poisson Hail on a Wireless Ground}

\author{F. Baccelli, 
K. Feng \ and S. Foss\thanks{F. Baccelli is with INRIA Paris and Telecom Paris.}\thanks{K. Feng started this work while with INRIA Paris and is now with CNRS-ETIS.}\thanks{S. Foss is with the School of MACS, Heriot-Watt University
and
Sobolev Institute of Mathematics.}}

\date{January 15, 2025}
\begin{document}

\maketitle
\begin{abstract}
    This paper defines a new model which incorporates three key ingredients
of a large class of wireless communication systems: (1) spatial interactions
through interference, (2) dynamics of the queueing type, with users joining and leaving,
and (3) carrier sensing and collision avoidance as used in, e.g., WiFi.
In systems using (3), rather than directly accessing the shared resources upon arrival,
a customer is considerate and waits to access them until nearby users in service have left.
This new model can be seen as a missing
piece of a larger puzzle that contains such dynamics as spatial birth-and-death processes, the Poisson-Hail model, and wireless dynamics as key other pieces.
It is shown that, under natural assumptions, this model can be represented as a
Markov process on the space of counting measures.
The main results are then two-fold. The first result is on the shape of the stability
region and, more precisely, on the characterization of the
critical value of the arrival rate that separates stability from instability.
The second result is of a more qualitative or perhaps even ethical nature.
There is evidence that for natural values of the system parameters,
{{the implementation of sensing and the delayed access for collision avoidance can stabilize
a system}} that would be unstable if immediate access to the shared
resources would be granted. In other words, for these parameters, renouncing greedy access
makes sharing sustainable, whereas indulging in greedy access kills the system.

\end{abstract}
\input{Taxonomy_and_Motivations}
\input{Contributions}
\input{Description_of_Dynamics}

\input{Stability}
\input{Simulations}
\input{Spatial_Interacting_Queuing_Processes}

\input{Discussion_and_Open_Questions}
\input{appendix}

\bibliography{ref}
\bibliographystyle{ieeetr}

\end{document}

%% file: Taxonomy_and_Motivations.tex
%
%

\section{Introduction}
\subsection{Background}
\label{sec: Taxonomy and Motivations}
Spatial regulation is a central issue in the access to the wireless network medium.
It is somewhat surprising that such a regulation is fundamental in carrier-sense multiple access with collision avoidance (CSMA-CA) protocols, for instance,
where close-by simultaneous access (``collision") to the shared medium is prevented (``avoided"),
whereas it is only present through optional spatial reuse mechanisms in most
3GPP medium access schemes. The rationale for such a regulation is, of course, the mitigation
of interference: when treating interference as noise, a wireless system sees its performance
significantly degrade when the power of interference is high.
The general question is that of the optimal trade-off between the gains obtained on the communication
rates of scheduled users by delaying medium access of the other users,
and the negative consequences of the delays incurred by the latter.

This question has been central since the 70's. One of the difficulties is the definition of
metrics that incorporate in a rigorous way the spatial components that are inherent
in this class of problems and the dynamical system components of the problem. Among the latter,
we would quote the epochs at which packets to be transmitted by their wireless links
are generated by users, the delays which are incurred by these packets due to spatial regulation
and more generally due to local queuing, as well of course as delays which are incurred by them
for transmission on the wireless links once they are scheduled.

The present paper proposes a new model that is meant to address this old question in a new way.
This model integrates spatial arrival processes, spatial regulation as described above, as
well as an exact representation of scheduled wireless link transmission rates based on classical
information-theoretic paradigms. In contrast to earlier approaches, this new model gives an exact
representation of the resulting dynamics and is amenable to a mathematical analysis
that is not based on an approximation (there is no mean-field approximation in particular).
It allows us to define a rigorous metric, which is the dynamic stability capacity of the system.
This new model is used to prove that appropriate spatial regulation may allow
one to increase this dynamic stability capacity. This is the central result of the paper, together
with the definition of the setting allowing one to formalize the problem.

\subsection{Taxonomy and Motivations}
We discuss here four basic classes of spatial queuing dynamics.
In each case, there is a space-time Poisson process of customer arrivals on a compact ${\cal X}\subset \mathbb{R}^d$.
Arrivals are characterized by a time $t\in \mathbb R$ and a locus $x\in \cal X$.
Each arrival also comes with a service demand (or file size) $h\in {\mathbb R}^+$,
that may vary from customer to customer.
The compact $\cal X$ can be seen as the server.
{{
The intensity measure of this Poisson process is assumed to be of the form
$\lambda \mathrm{d}t m(\mathrm{d} x)$, where $\lambda>0$ and $m$ is a measure on $\mathcal X$ with $m(\cal X)<\infty$.}}

It is best to start with the {\em Poisson Hail on a Hot Ground} (PHHG) dynamics introduced in \cite{baccelli_foss_2011}.
In this model, in addition to their service demand, each arrival at locus $x$ comes
with an exclusion set $S\subset \mathbb{R}^d$. A typical instance, which is used in this introduction,
is that where $S = B(x,R)$, namely the closed ball with radius $R$ centered at $x$,
where $R>0$ is a fixed constant or a random variable.
Two arrivals are said to be conflicting if their exclusion sets intersect. No two conflicting
customers can be served simultaneously. The service of a customer with arrival time $t$
starts when all conflicting customers that arrived before $t$ have left the system.
This can be seen as a local First-Come-First-Served (FCFS) rule.
The service speed is constant. Here is what explains the terminology.
The ball of a customer {{can be seen as}} the footprint of the hailstone associated with this customer and its service demand is 
the height of this hailstone. The temperature of the compact $\cal X$ (subset of the ground) is constant
so that the hailstones that touch the ground melt at constant speed, whereas the others form a (cold) pile-shaped
spatial queue.

The second dynamics is that obtained from the PHHG when setting $S=\{x\}$ (or $R=0$),
namely a hailstone arriving at $x$ is an ice-stick at $x$. {{In this case, when the measure $m$ is absolutely
continuous w.r.t. Lebesgue measure}}, no two customers ever conflict a.s., so that each stick
receives immediate attention and melts at constant speed. This model is the spatial birth-and-death (SBD) process introduced by Preston in \cite{preston1975spatial}. 
If the space marginal is atomic, we get a queuing network consisting of a finite collection
of independent FCFS M/M/1 queues located {{at the atoms of $m$}} 
if the height variables are exponential and i.i.d.

The third model is the wireless spatial birth and death (WSBD) process which is a variant of 
the SBD introduced in \cite{sankararaman2017spatial}.
The only difference from the SBD is in the service rate (or melting speed) of the sticks.
Rather than a constant speed, the service rate of a customer at $x$ 
is proportional to the Shannon rate at $x$ when considering
all other concurrent customers as interferers and treating interference as noise. 
The rationale for this service rate comes from information theory and wireless communications \cite{tse2005fundamentals}.
The service discipline is hence with immediate attention but with varying service rates across time and space.

The present paper is focused on a model that is new, to the best of our knowledge, and that we will refer to as the Poisson hail
on a Wireless Ground (PHWG) model. It is meant to study arrivals with (non-trivial) exclusion sets and varying
service rates across time and space. In the wireless setting, the exclusion set serves as a protection
zone around wireless transmissions, which is a way to control the medium access similar to what is introduced
in carrier-sense multiple access protocols (see below). The dynamics can be described 
in more detail as follows:
\begin{itemize}
\item  {\bf{Local FCFS discipline:}} A hailstone only starts being served (being active)
when all conflicting hailstones that precede it have left.
\item  {\bf{Shannon rate:}} The service rate of each active hailstone at any time $t$ is its Shannon rate,
which is a deterministic function of the configuration of all active hailstones at time $t$ (see Eq (\ref{eq1})). Note that, at a given time $t$, this rate may be different for two customers being served at that time.
\item  {\bf{Departure at service completion:}} As soon as a hailstone has fully melted
(its service demand is exhausted), it leaves the system.
\end{itemize}
This model can be seen as a variant of previous models in two equivalent ways: 
\begin{itemize}
\item A variant of the PHHG model when the temperature of the ground varies
with \textit{time and space} according to Shannon's capacity formula, namely the melting speed at time $t$ of 
a hailstone touching the ground and centered at $x$ is given by the Shannon rate,
which depends on the locations of all hailstones touching the ground at that time.
\item A variant of the WSBD model where, rather than starting transmitting immediately upon arrival,
an arrival is required to apply certain priority rules based on their exclusion sets and which are those of the PHHG model.
\end{itemize}

The rationale for introducing this model goes beyond the sake of completeness.
As suggested by (parts of) the terminology, our practical motivations are to be found in
the analysis of wireless communications (see Section \ref{sec: Contributions}) at the system level.
System-level properties include, for instance, the capacity region of the queuing system
or the mean system time of {the typical customer}.
By capacity, we mean the queuing theoretic capacity, namely the set of values for $\lambda$
such that the dynamics in question is stable (say positive recurrent under a proper Markov representation), when
fixing all other system parameters. The fact that this stability region
of PHWG is of the form $\lambda<\lambda_c$ when all other parameters are fixed is one
of the main results established in the present paper, together with a constructive representation of $\lambda_c$.
As we shall see, there are system parameter regions for which the WSBD is unstable, whereas the PHWG is stable.
In more concrete terms, in the stability region in question, the queuing delays incurred by some customers
due to the exclusion/protection rule, in fact, allow one to get finite system times in the steady state,
whereas stationary system times are infinite in the situation with immediate access to the channel ($R=0$).
We find it interesting that refraining from immediate access to a system could stabilize the system in question. 
This can be seen as a strong argument in favor of protocols of the CSMA-CA-type (see Section \ref{sec: Contributions}),
which control the access to the shared medium, in order to prevent nearby customers from accessing that shared channel at the same time.
In the stable case, we also discuss the stationary distribution of the waiting
or system time of the typical customer and its dependence on $R$.
Even if wireless motivations are essential, we will show (see Section \ref{sec: SIBD}) that this class of models
has, in fact, a much broader scope and potential set of applications.


%% file: Contributions.tex
\subsection{Contributions}
\label{sec: Contributions}

The SBD dynamics was first studied in \cite{preston1975spatial} and was then the subject of numerous further studies.
The PHHG dynamics was introduced in \cite{baccelli_foss_2011}. The stability condition is known. The steady state
can be constructed for both compact and infinite spaces. Both dynamics are well understood.
The WSBD model was introduced in \cite{sankararaman2017spatial} for the compact and continuous space case. 
A model with the infinite discrete space case was analyzed in \cite{sankararaman2019interference}, where the stability conditions were obtained and a closed form was derived for the steady
state queue size. The infinite continuous state space case
is unsolved at this stage, to the best of our knowledge.

As explained above, the PHWG model, which is the central topic of the present paper, admits both
WSBD and PHHG (and hence SBD) as special cases.
In view of the state of the art for these special cases,
we will restrict ourselves to the compact state space case. 
The first contributions of the paper are the construction of the steady state and the identification of the stability region of PHWG.
{Our main result on this matter is presented in Theorems 1 and 2: we prove that, when fixing
all parameters of the model except $\lambda$, there exists a critical
value $\lambda_c$ such that if $\lambda < \lambda_c$, the PHWG dynamics is a
positive recurrent continuous Markov chain in the space of counting measures,
while it is transient if $\lambda > \lambda_c$.
Another contribution is the comparison of PHWG with the dynamics listed above.
We study two examples in Section \ref{sec: Simulations}, one in the continuous-space setting and one in the discrete-space setting, to show that
PHWG outperforms WSBD for natural values of the
system parameters. Since WSBD is obtained when setting $R=0$ in PHWG,
a simple instance is the identification of an optimum of $\lambda_c$ for $R=R^*$ with $R^*\ne 0$, and the determination of $R^*$ through numerical evaluations.}
A third outcome of the paper is the identification of
extensions of this model to spatial birth and death with interaction (or death) functions
other than that associated with the Shannon rate. We show that there is a wide class of dynamics 
where the approach and results developed here extend in a natural way.

\subsubsection{Mathematical Approach}
A key step in the construction of the steady state and the identification of the 
stability region is the identification of arrival blocks that form i.i.d. sequences.
A sufficient condition to construct such blocks is the existence of what we call
 {\em{zigzag}} events. The simplest such event consists of a
single exclusion set that covers the entire space. As we show below, more
refined zigzag events occur whenever each exclusion set contains a ball of 
deterministic radius $R$ where $R$ is any positive constant common to all arrivals.
More refined events can be considered, linked to the rank of tropical matrices.

\subsubsection{Implications for Wireless Networks}
The main novelty of the paper in terms of wireless communications is in the analysis of the
dynamics of the spatial properties of carrier-sense multiple access with collision avoidance (CSMA/CA).
The principle of CSMA/CA consists in sensing the shared medium and refraining from accessing
it if there is a nearby transmitter already using it.
There is a great deal of literature on this multiple access scheme. 
Its spatial but static analysis was first considered using a Matérn  point 
process heuristic in \cite{nguyen}, without analysis of dynamic stability. \cite{chine} defines and analyzes the conditions for $\epsilon$-stability, where at most $\epsilon$-fraction of wireless queues are instable.
Mean-field heuristics associated with the dynamics were proposed 
\cite{grec}, with an approach ignoring the spatial nature of the problem and in \cite{italien} when
taking this feature into account.
To the best of our knowledge, the present paper 
is the first to propose an {\em exact} (no mean-field approximation) analysis of the spatial properties of CSMA/CA type dynamics.

It is important to stress that the version of the collision avoidance mechanism studied in the
present paper is quite different from that used in the current versions of
the CSMA-CA protocol. The main common feature is the implementation of an exclusion
principle which forbids that two (or more) customers with a mutual distance
smaller than a given value could transmit simultaneously.
The local FCFS mechanism used in the paper in order to implement this
is quite different from what is currently done.

\subsection{Paper Organization}
The paper is structured as follows:
Section \ref{sec: Description of Dynamics} defines the dynamics.
Section \ref{sec: Stability} establishes the main stability result.
Section \ref{sec: Simulations} gathers simulation and comparison results.
Section \ref{sec: SIBD} gives the class of Spatial Interaction Birth and Death (SIBD) processes
to which our main results can be extended.
Section \ref{sec:open} contains various discussions and a list of open problems.
Finally, the appendix gathers the proofs of our two main theorems.



%% file: Description_of_Dynamics.tex
\section{Description of the Dynamics}
\label{sec: Description of Dynamics}
	The aim of this section is to show that the sample path of the random dynamics described
	informally in the last section can be constructed by induction. 
    The main outcome of the construction is
	that there are no accumulation points of jumps in the dynamics so that the latter is well defined for all positive times for any given initial condition.

\subsection{Arrival Process}
\label{ssap}
{{
The arrival process is best described 
as a stationary marked Poisson point process on
$\mathbb R$ with i.i.d. marks on $\mathcal X\times \mathbb{R}^2$.  The
intensity of the arrival process is denoted by $\Lambda >0$.
The probability distribution of the marks is denoted by $\mu\times \nu$,
where $\mu$ is a probability measure on $\mathcal X$ and $\nu$ is a probability
distribution on $\mathbb{R}^2$. This marked point process is denoted by
$\mathcal{A} = \{\left((t_i),(x_i,h_i,R_i)\right)\}_{i\in\mathbb{N}^+}$,
where $t_i$ denotes the arrival time of the $i$-th customer 
(arrivals are ordered by increasing arrival times), $x_i$
the arrival locus of the $i$-th customer, $h_i$ the service time 
of the $i$-th customer, and $R_i$ its exclusion radius. 
The interpretation is that the $i$-th arrival
comes as a ``hailstone'' $(S_i,h_i)$, where $S_i=B(x_i,R_i)$ is the
``footprint'' (which is here closed ball of radius) 
and $h_i\in\mathbb{R}^+$ is the ``height'' of the hailstone.
The random sequence of pairs $\{(h_i,R_i)\}$ is assumed to be i.i.d. and not depend on 
$\{(x_i,t_i)\}$. However,
$R_i$ and $h_i$ need not be independent. They are assumed to have a joint distribution
$\nu$ on $\mathbb{R}^2$. 
This formulation is equivalent to the informal models of the introduction with
\begin{equation}\Lambda=\lambda m(\mathcal{X}) \quad \mbox{and} \quad \mu(\cdot) = \frac{m(\cdot)}{m(\mathcal{X})}.\end{equation}
In what follows, we will often omit the index $i$ if it is not needed.
More precisely, $(x,h,R)$ will denote an independent copy of $(x_1,h_1,R_1)$
and $S$ will denote the set $S=B(x,R)$.}}
In the wireless setting, the $i$-th customer is a wireless link where the transmitter and
receiver are co-located. This simplification makes sense when the distance from
transmitter to receiver is small compared to the distances between links. 

\begin{figure}
    \centering
\includegraphics[width=0.5\linewidth]{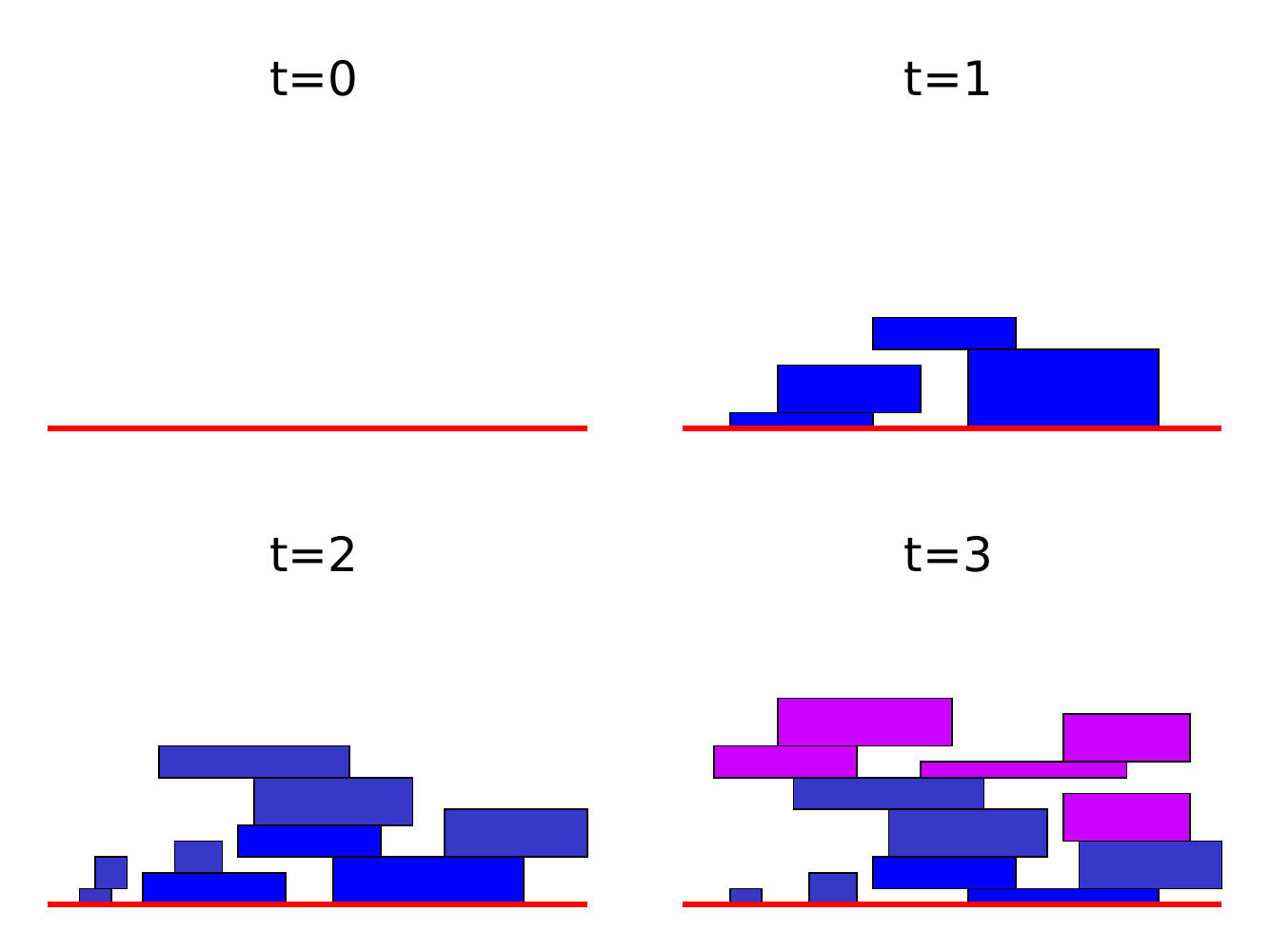}
    \caption{An illustration of the dynamics on $\cal X \subset \mathbb{R}$. The hot ground (or equivalently, $\cal X$) is illustrated by the red line segment. Each rectangle represents an arrival, where the height represents the {(residual)} file size, and the projection of the rectangle on the ground represents the exclusion set. }
    \label{fig:chang}
\end{figure}

The dynamics is depicted in Figure \ref{fig:chang}. 

\subsection{Service Discipline and Service Rate}

The service discipline is locally FCFS:
the service of customer $n$ (arriving at time $t_n$ at $x_n$) starts as soon as all customers $k$
arriving before $t_n$ and such that $S_n\cap S_k\ne\emptyset$ have left.

Throughout the paper, an attenuation function $l$ is assumed to be given. This is a function from $\mathbb R^+$ to $\mathbb R^+$, which
is assumed to be non-increasing and bounded from above. The positive constants
$b,s$, and $w$ represent the bandwidth,
the signal power and the additive white Gaussian noise power, respectively. 
With this notation, the service rate of customer $n$ at any time $t$, when it is active and still in the system, is the Shannon rate
\begin{equation}
\label{eq1}
b\log\left(1+\frac{s}{w+\sum_{x_k\in\phi_t\setminus\{x_n\}} l(||x_k-x_n||)}\right),
\end{equation}
with $\phi_t$ the point process made of all points $x_k$ of customers still in the system and
being currently served at time $t$. $\|\cdot\|$ denotes the Euclidean metric of $\cdot\in\mathbb{R}^d$. We make  two observations in the following regarding the service rate.

\subsubsection{Upper Bound on Service Rate}
Shannon rates are bounded from above in the sense that for all spatial configurations, the service rate of any customer is upper bounded by the constant $c:=b\log\left(1+s /w\right)$, which is the rate in the absence of interference.

\subsubsection{Lower Bound on Sum Service Rate}
One can give a positive constant lower bound to the sum of the
departure rates of all customers present in the system at any time, given that there is at least one
customer, as shown in the next lemma.
\begin{Lem}
\label{lemma1}
 Under the monotonicity and boundedness assumptions on the attenuation function, if there is at least one active customer
 in the system, then the sum rate of all active customers in the system
 is bounded below by a positive constant.
\end{Lem}
\begin{proof}
Assuming that there are $N>0$ active customers in the system, the sum rate is
\begin{eqnarray*}
	& & \hspace{-1cm}	\sum_{n=1}^N b\log\left(1+\frac{s}{w+\sum_{x_k\in\phi_t\setminus\{x_n\}} l(||x_k-x_n||)}\right) \\
	&\pgeq{a} &b
	\log\left(1+
	\sum_{n=1}^N
	\frac{s}{w+\sum_{x_k\in\phi_t\setminus\{x_n\}} l(||x_k-x_n||)}\right) \\
	&\pgeq{b} &
	b
	\log\left(1+\frac{Ns}{w+ (N-1) L}\right),
\end{eqnarray*}
where $L$ is an upper bound on $l$. Step (a) follows from the inequality $\prod_{n=1}^N(1+a_n)\geq 1+\sum_{n=1}^N a_n$ for $a_n\geq0$. Step (b) follows from the upper bound on $l$.
The fact that the function $N\to {Ns}/{(w+(N-1)L)}$ is uniformly bounded below concludes the proof.
\end{proof}

\subsection{Construction of the Model}
{{We construct the model for ${\cal X}\subset\mathbb{R}^d$, with Poisson arrivals with intensity $\Lambda$. We assume here that $\nu$ has its first marginal $F$ with finite mean. So, the
service demands (file sizes) are i.i.d. with distribution $F$.}}
We define a sequence of directed acyclic graphs (DAGs)
\footnote{A DAG is a directed graph with no directed cycles.}, denoted by $\{G_n\}$, $n=0,1,2,\ldots,$
and an increasing sequence of times $\{j_n\}$ by induction w.r.t. $n$.
The continuous-time process that we build is piecewise constant with
jump times $\{j_n\}$, where $j_n$ is a strictly increasing sequence.
DAG $G_n$ is the state DAG on the time interval $[j_n,j_{n+1})$.
A DAG is possibly non-connected. DAG $G_n$ has a set of vertices (customers) $V_n$
with acyclic directed edges $E_n$ between them.
The service (or active) set $U_n$ of $G_n$ is the subset of vertices of $V_n$
without predecessor\footnote{Vertex $v$ is a predecessor of $u$ if there exists a directed edge from $v$ to $u$. }.
Each node $v\in V_n$ has a location $x(v)\in {\cal X}$, an exclusion set $S(v)$,
a birth time $b(v)$, and a death time $d(v)$. The vertices of $G_n$ have each an additional state variable
which is its residual service demand (file size) at time $j_n$, denoted by $z_n(v)\in {\mathbb{R}}^+$.

In order to define the dynamics, we need

{{
\begin{itemize}
	\item The sequence of i.i.d. exponential random variables $\{\widetilde \tau_n\}$ of parameter $\Lambda$; $\widetilde \tau_n$ represents the time that elapses between $j_n$ and the next {\em{potential}} arrival after time $j_n$ (by ``potential", we mean the following: if $\widetilde \tau_n\ge q_n$ defined below, then $\widetilde \tau_n$ is not used and at time $j_{n+1}$, a new exponential random variable $\widetilde \tau_{n+1}$ is sampled to locate the next potential arrival epoch seen from time $j_{n+1}$. This representation
    is fine because of the memoryless property of exponential random variables. If 
    $\widetilde \tau_n<q_n$, then the next arrival takes place at time $j_n+\widetilde\tau_n$
    and then, the random variable $\widetilde \tau_n$ is used ``effectively").
    \item an i.i.d. sequence of uniform random variables $\{\widetilde X_n\}$ on $\cal{X}$;
		$\widetilde X_{n+1}$ represents the locus of the potential next arrival after time $j_n$.
	\item an i.i.d. sequence of pairs of random variables $\{(\widetilde R_n, \widetilde h_n)\}$, where 
		$\widetilde S_{n+1} = B(\widetilde X_{n+1},\widetilde R_{n+1})$ represents the footprint of the  potential next arrival after time $j_n$, and $\widetilde h_{n+1}$ its file size.
\end{itemize}
All the sequences of random variables above are mutually independent, while dependence between $\widetilde R_i$ and $\widetilde h_i$ is allowed.
}}

The word ``potential'' here means that this variable can be used or discarded depending on the conditions
spelled out in the induction. The tilda is used to distinguish it from the effective variable. The induction is as follows.
When the DAG at $j_n$ is $G_n$, define the potential departure time of node $v\in U_n$ as
$$p_n(v)=z_n(v)/r_n(v),$$
with $r_n(v)$ the Shannon rate of $v$ in the spatial configuration 
\begin{equation} \Psi_n:=\{{x(v) \colon v\in  U_n} \},\end{equation}
that is
\begin{equation}
	\label{eq:sh}
	r_n(v)= b \log\left(1+ \frac{s}{I_n(v)+w}\right),
\end{equation}
with 
\begin{equation} 
I_n(v)= \sum_{x(u)\in \Psi_n\setminus\{x(v)\}} l(||x(v)-x(u)||)\end{equation}
the total interference power seen at  $v$. Note that $\Psi_n = \phi_{j_n}$ (previously defined in Eq (\ref{eq1})).
	Let
\begin{equation}q_n :=\min_{v\in U_n} p_n(v)\end{equation}
denote the time elapsed until the next departure given the DAG at $j_n$.

\begin{itemize}
	\item If $q_n > \widetilde \tau_n$, then
		\begin{itemize}
		      \item $j_{n+1}:=j_n +\widetilde \tau_n$.
		      \item $V_{n+1}:= V_n\cup \{v'\}$ where 
		              the extra node $v'$ is a new arrival, with an assigned location $\widetilde X_{n+1}$, an
				exclusion set $\widetilde S_{n+1}$,
				and residual service time $\widetilde h_{n+1}$, i.e., 
				$x(v')=\widetilde X_{n+1},~b(v') = j_{n+1},~S(v')= \widetilde S_{n+1},~z(v') = \widetilde h_{n+1}$.
			\item $E_{n+1}$ is the union of $E_n$ and the set of edges from $V_n$ to ${v'}$, defined as
				follows: there is an edge from $v\in V_n$ to ${v'}$ if $S(v')\cap S(v)\ne \emptyset$ (to enforce the local FCFS policy).
				This defines $G_{n+1}$.
		      \item For all $v\in U_n$, $x(v)$ is unchanged and the residual file size is updated to 
			      $z_{n+1}(v)=z_n(v)-\widetilde \tau_n r_n(v)$.
		      \item If ${v'}$ has no predecessor in $G_{n+1}$,
			      then $U_{n+1}:=U_n\cup\{{v'}\}$.
			      Otherwise, $U_{n+1}:=U_n$.
		\end{itemize}
	\item If $q_n {\leq} \widetilde \tau_n$, then, with $u$ denoting the argmin in the definition of $q_n$, 
		\begin{itemize}
		      \item $j_{n+1}:=j_n +q_n$.
		      \item $V_{n+1}:= V_n\setminus \{u\}$  (departure of customer $u$), $d(u) = j_{n+1}$.
		      \item $E_{n+1}$ is the restriction of $E_n$ to $V_{n+1}$.
              \item $U_{n+1}$ is the set of vertices without predecessor in $(V_{n+1},E_{n+1})$.
				This defines $G_{n+1}$.
		      \item For all $v\in U_{n+1}$, the residual service time is updated to
			      $z_{n+1}(v)=z_n(v)-q_n r_n(v)$.
		      
		\end{itemize}
\end{itemize}
This induction step defines $j_{n+1}$, $G_{n+1}$ and the residual service times at time $j_{n+1}$.
The initial DAG $G_0$ is, for instance, the empty DAG and $j_0=0$.
Up to ties, which occur with 0 probability, the above dynamics defines a
piecewise constant DAG state process up to time $j=\lim_{n\to \infty} j_n$.

Some basic observations are in order with respect to the model and its construction.
\begin{enumerate}
\item If the initial condition is a finite DAG, then
$j=\infty$ a.s. In the empty graph initial condition case, this follows from the fact that for any $k>0$, at the arrival
time $t_k$ of the $k$-th customer, at most $k$ departures can have taken place. Since $t_k$ tends to
infinity a.s. as $k\to \infty$, we obtain that no finite accumulation points can exist for the
sequence $\{j_n\}_n$ a.s. This can directly be extended to the case where the initial condition is any
finite DAG.

\item Each DAG can be represented as a marked counting measure
where the atoms of the counting measure are the locations of the
customers present in the system and the marks are the precedence edges and the residual service times
of these customers (all those having a predecessor in the DAG have their residual
file size equal to their full initial file size).



\item 
{In the special case where ${\cal X} = [-Q,Q]^d$ and all exclusion sets contain a {ball of fixed radius $\overline R>0$,}} 
no two customers with intersecting exclusion sets are served at the same time, so that
the point process of active customers (at any time) is a hardcore point process
with balls of radius $\overline R>0$. 
This implies that the total number of customers that are
served at any given time is bounded above by a constant {{that depends on $Q$ and $\overline R$.}}
This has several implications. 
First, even if the initial DAG is infinite, we still have $j=\infty$ a.s. in the last construction algorithm.
Indeed, we have at most a certain number of customers being served without interruption of service, each with a rate bounded
from above by $c$ (see the previous subsection),
and it is easy to see that this implies that there can be no finite accumulation point 
of departures a.s. under the foregoing Poisson assumptions.
Second, since the function $l$ is bounded above, this also implies that
the interference is bounded above by a constant at any given time and locus.
Hence, the Shannon rate for any active customer is also bounded from
below by a positive constant at any given time \cite{feng2023spatial}.

\end{enumerate}

\subsection{Generalizations}
The previous construction leverages the memoryless property of the exponential
distribution of the interarrival time.  It can easily be extended to the case where the arrival process
is a renewal process without multiplicity (rather than a Poisson point process) where the inter-arrival times $t_{i+1}-t_i$ are assumed to be i.i.d. and strictly positive with a finite mean,
by adding in the induction a supplementary state variable for the remaining time to the next arrival.  
All the steps and results of the last subsection then hold for this case.
{{It is also easy to extend this framework to the case where footprints
are random compact sets that contain balls. }}

%% file: Stability.tex
\section{Stability}
\label{sec: Stability}
The aim of this section is to establish the main stability result of the paper.
This is done by reduction of the dynamics described in the last sections to a Lindley type
recurrence equation. This reduction relies on assumptions on the arrival process
and the exclusion sets which we call zigzag events. Under this assumption, the
arrival process can be decomposed into i.i.d. blocks such that the waiting time
of the first customer in a block can be determined from that of the last
customer in the preceding block. This leads to the identification of an embedded
Markov chain on the positive real line, whose positive recurrence can be assessed
in a direct way by some natural Lyapunov function argument.
The section adopts the assumptions of Section \ref{sec: Description of Dynamics}.A and Section \ref{sec: Description of Dynamics}.B. It makes further assumptions on the structure of the exclusion sets.

\subsection{Main Statistical Assumptions}
\label{ss:maes}
\subsubsection{Assumptions on the Arrival Process}
\label{sss:msa} 
We recall here for easy access the following 
default assumptions for this section:

\noindent 
{{{\bf Assumption A1.} The arrival process $\sum \delta_{t_i}$ is
Poisson with intensity $\Lambda$ and i.i.d. marks with probability distribution
$\mu \times \nu$  on 
${\mathcal X}\times \mathbb{R}^2$.\\

\noindent
{\bf Assumption A2}. The marks $(R_i,h_i)$ (which are i.i.d.) are such that
the $h_i$ random variables have finite mean.\\

\subsubsection{Assumptions on Exclusion Sets}
We present here our third set of assumptions:\\

\noindent
{\bf Assumption A3.} 
There exists a positive integer $k$ and a sequence of non-empty deterministic
measurable sets $B_1,B_2, \ldots,B_k$ in the space ${\cal X}$ such that
\begin{itemize}
\item{(a)}
$B_{i+1}\cap B_{i} \neq \emptyset$, for any $i=1,2,\ldots,k-1$;
\item{(b)}
$B_{i+1} \setminus \cup_{j=1}^i B_j \neq \emptyset$, for any $i=1,2,\ldots,k-1$;
\item{(c)}
$\cup_{i=1}^k B_i = {\cal X}$;
\item{(d)} 
${\mathbb P} (B_i \subseteq S)>0$, for all $i=1,2,\ldots,k$, where $S=B(x,R)$ (see 
Subsection \ref{ssap}).
\end{itemize}}}


\begin{Rem}
Note that conditions (a)-(c) are always satisfied (for instance, with $k=1$ and
$B_1={\cal X}$), so that the key ingredient in this assumption is (d) in fact. For example,
in the special case of $S=\{x\}$ a.s. for an arrival at $x$, the assumptions cannot be jointly satisfied for $|{\mathcal X}|\neq 0$.
\end{Rem}

\begin{Rem}
{{In the special case where every exclusion set contains a ball of fixed radius $\overline R>0$ centered
on the arrival point, it is easy to see that the last property holds for an integer $k$ large enough,
which is linked to the number of balls of radius $\overline R$ needed to completely cover the window of interest $\cal X$.}}
However, there may exist a collection of subsets that satisfy the same property for a smaller $k$. For example,
the last assumptions hold for $k=1$ when there is a positive probability that an arrival
comes with $S = \cal X$. This is the case for i.i.d. exclusion radii with exponential distribution. 
\end{Rem}

\begin{figure}
    \centering
        \includegraphics[trim={17cm 6cm 13cm 5cm},clip,width=0.485\linewidth]{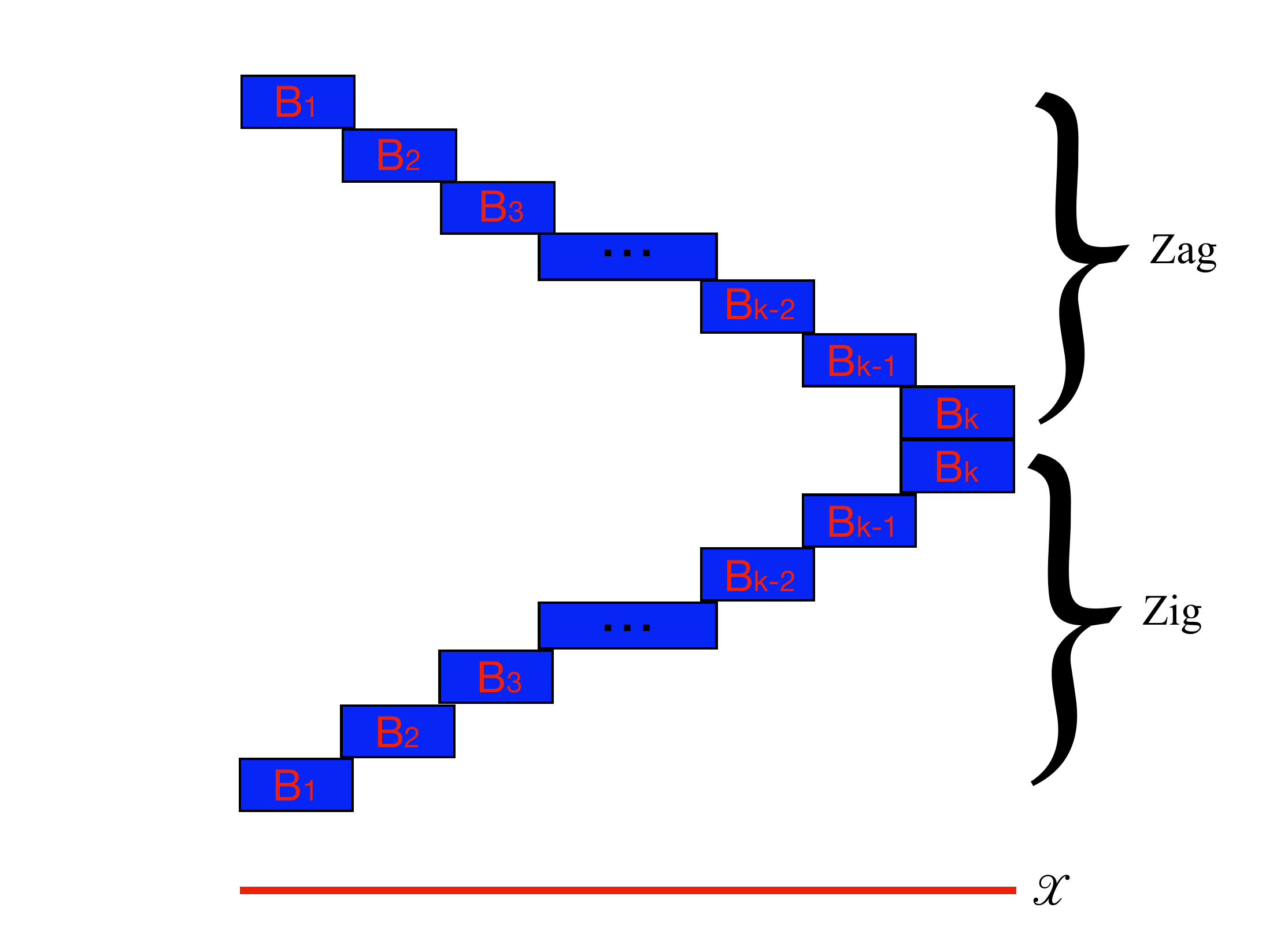}
    \caption{An illustration of an underlying deterministic sequence $\{B_i\}_{i=1}^k$ with the corresponding zigzag structure for $\cal X \subset\mathbb{R}$.}
    \label{fig:zigzag}
\end{figure}
For all the following results, Assumptions 1-3 are assumed to hold.

\subsection{Block Representation}
\label{ss:br}
Consider the events $A_{i,n} = \{ B_i \subseteq S_n\}$, $n\ge 1$, and, for all $n\geq k+1$, the events
\begin{align}
\label{eq:cn}
C_{n} =\left(\bigcap_{i=0}^{k-1} A_{i+1,n-k+i} \right)
\bigcap 
\left(\bigcap_{j=0}^{k-1} A_{k-j,n+j}\right)
\equiv C_{n,1} \cap C_{n,2}.
\end{align}
In what follows, we will call $C_n$ a {\em zigzag event}. We refer to $n$ as the 
``center" of this event $C_n$.  It is composed of the ``zig'' event $C_{n,1}$, of ``length" $k$ (i.e. composed of $k$ events), and then of the ``zag" event
and $C_{n,2}$, also of length $k$. The length of the zigzag event is hence $2k$.
Note that, for any $m$ and for any $1\leq i_1,\ldots,i_m\leq k$ and $n_1<n_2<\ldots < n_m$, the events $(A_{i_j,n_j})_{1\leq j\leq m}$ are mutually independent. Therefore, for any $n$, the events $C_{n,1}$ and $C_{n,2}$ are mutually independent and occur with the same (positive) probability.

Fig. \ref{fig:zigzag} illustrates a sequence $\{B_i\}_{i=1}^k$ that forms a zigzag event for $\cal X \subset \mathbb{R}$ satisfying assumptions (a)-(c).

A zigzag event creates an isolation between customers who arrive before the zig and customers who arrive after the zag, where the latter are served only after the former have left the system. Based on zigzag events, we now introduce some splitting of customers into ``blocks", or ``cycles". Let $n_0=0$, $n_1 = 2k \cdot \min \{r\ge 1: {\mathbf I}(C_{2kr})=1\}$ and, for $m=1,2,\ldots$,
 \begin{align}\label{nm}
 n_{m+1} = n_m + 2k \cdot \min \{r \ge 1: {\mathbf I}(C_{n_m+2kr})=1\},
 \end{align}
  where ${\mathbf I}$ is the indicator function, which is equal to 1 if the event occurs and 0 otherwise. In words, $n_m$ is the center of the $m$-th zigzag event on the integer lattice of multiples of $2k$.

\begin{Def}
\label{def: block}
For $m\ge 1$, the customers numbered $n_{m}, n_{m}+1, \ldots, n_{m+1}-1$ form the $m$-th {\em block of customers}.
\begin{itemize}
\item The first customer of this block (that having an exclusion set containing $B_k$)
is the {\em opening customer} of block $m$; it arrives at $T_m$.
\item The last customer of this block (which also has an exclusion set containing $B_k$) is the {\em closing customer}
of block $m$; it arrives at $\widehat T_{m}:=t_{n_{m+1}-1}$.
\end{itemize}
\end{Def}

{For $m\geq0$, let \begin{align*}
\beta_m := n_{m+1}-n_{m},
\end{align*}
which denotes the number of customers in block $m$. The distribution of $\{\beta_m\}_{m\geq0}$ is characterized by the lemma below.

\begin{Lem}
\label{lem: geometric}
$\beta_0,\beta_1,...$ are i.i.d. and geometrically distributed  random variables with parameter $p_b$ multiplied by $2k$, where 
\begin{align*}
  p_b:= {\mathbb P} (C_n)  = \left( \prod_{i=1}^k {\mathbb P} (B_i \subseteq S) \right)^2 >0.
\end{align*}

 \end{Lem}
 \begin{proof}
   {Recall that the exclusion sets $\{S_i\}$ are independent. The events $\{C_{2k}, C_{4k}, C_{6k},...\}$ 
   are stationary and mutually
independent, since each of these events depends on exactly $2k$ exclusion sets with
indices drawn from disjoint sets.} Let $r_1=\min \{r\ge 1: {\mathbf I}(C_{2kr})=1\}$.
     $r_1$ is the index of the first success in Bernoulli trials ($C_{2kr_1}$ is the first zigzag event) with parameter $p_b$ and is geometrically distributed. The fact that $(n_{m+1}-n_m)/(2k)=((n_{m+1}+k-1)-(n_m+k-1))/(2k)$ is i.i.d. and geometrically distributed follows from the definition in Eq (\ref{nm}) and a similar argument as that of $r_1$. 
 \end{proof}
 
Lemma \ref{lem: geometric} shows that the random variables $\beta_0, \beta_1,\ldots$ and $n_1, n_2,\ldots$
are all a.s. finite. Specifically, ${\mathbb E}\beta_m = 2k/p_b.$}

Let $T_0=0$ and $T_m :=t_{n_m}$. For $m\ge 0$, let 
 \begin{align*}
	 Y_m :=\{((x_n,t_{n+1}-t_{n_m}), (S_n,h_n)), n=n_m, n_m+1,n_m+2,\ldots, n_{m+1}-1\}.
 \end{align*}

\begin{figure}
    \centering
    \subfigure[An illustration of a block of arrivals, $k=4$.]
    {
        \includegraphics[trim={17cm 6cm 13cm 5cm},clip,width=0.49\linewidth]{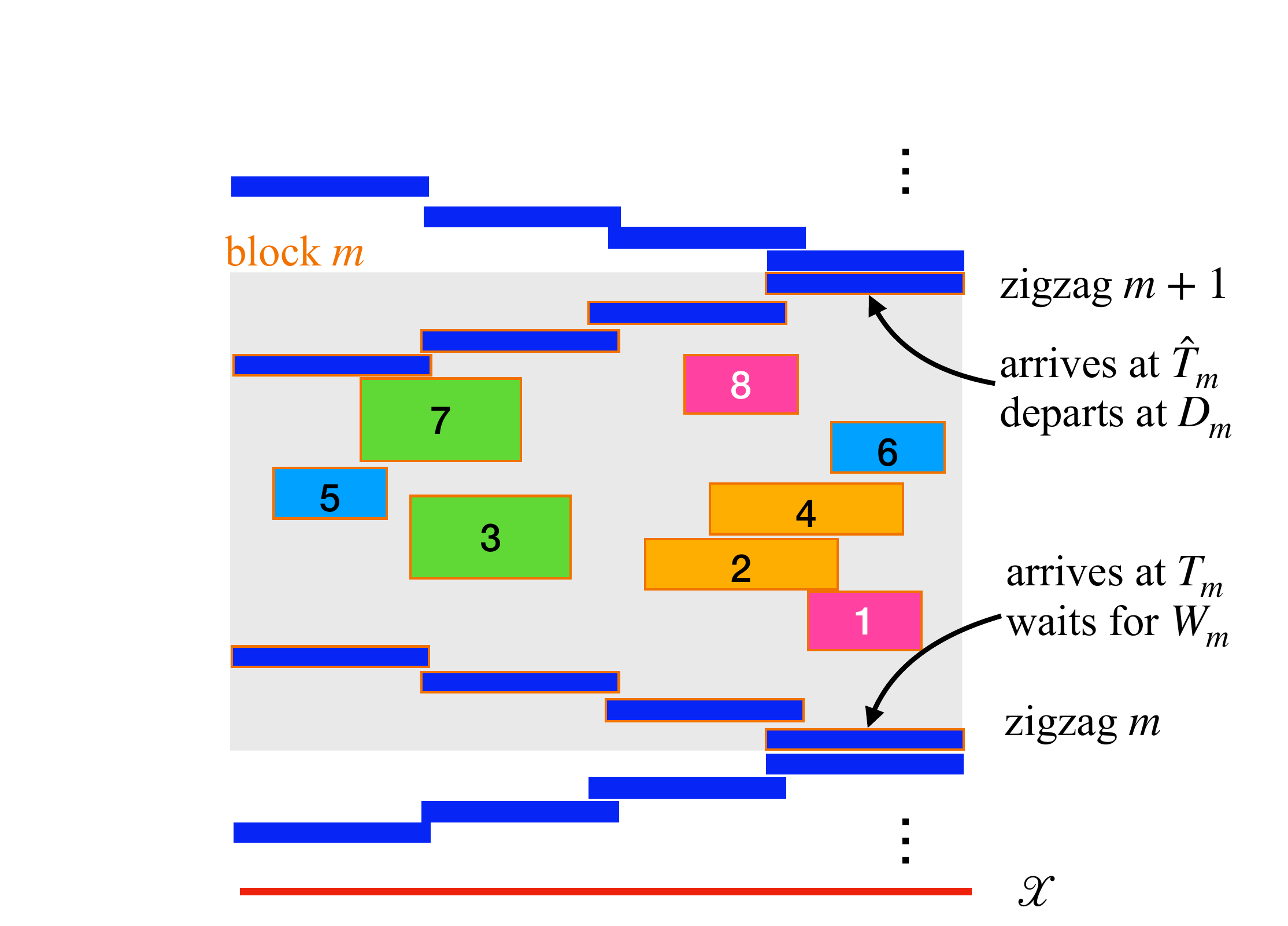}
        \label{fig:first_sub}
    }
    \subfigure[An equivalent representation of  block $m$ viewed at the beginning of its service when waiting time $W_m\geq\widehat{T}_m-T_m$.]
    {
        \includegraphics[trim={18cm 6cm 18cm 10cm},clip,width=0.47\linewidth]{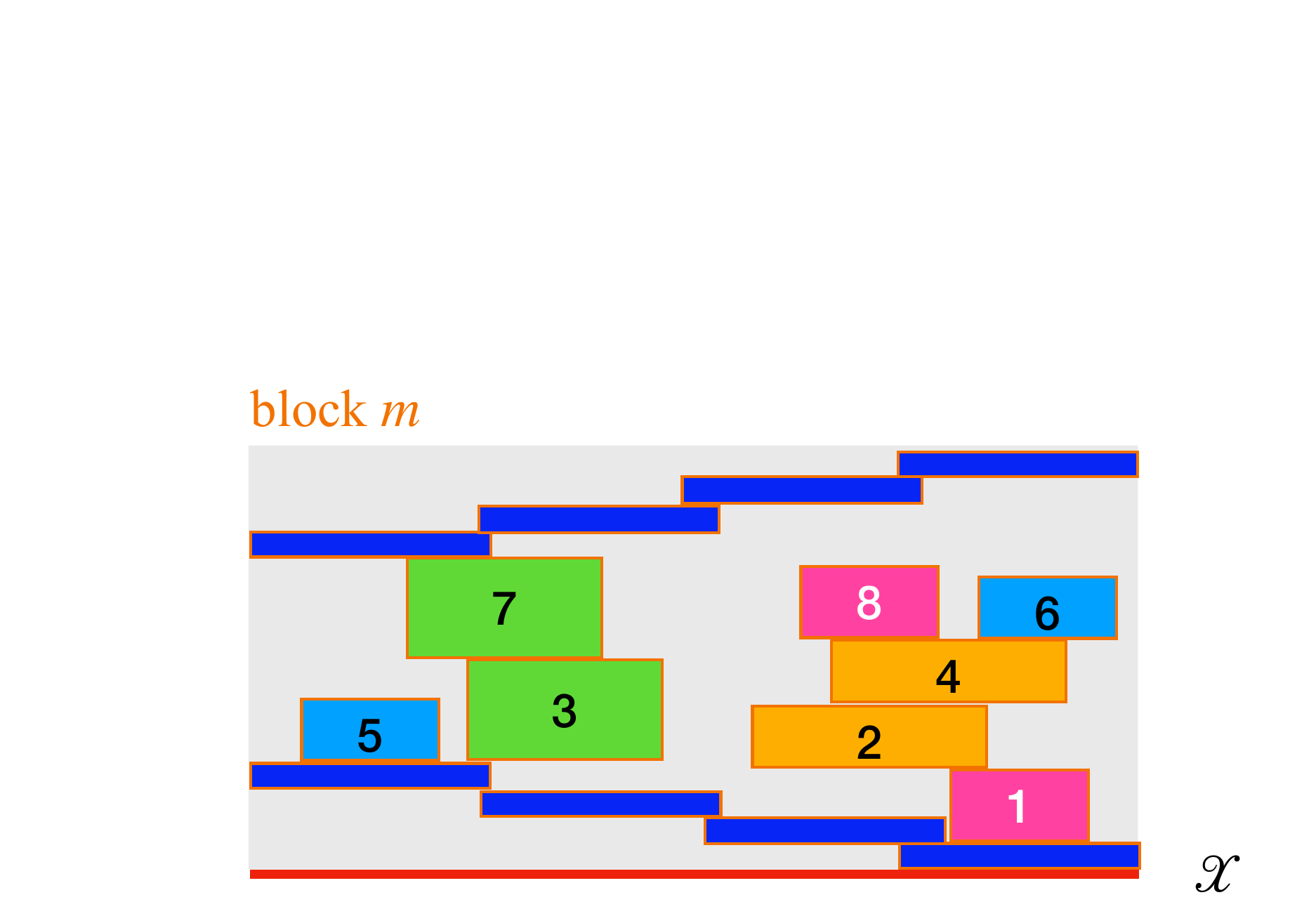}
        \label{fig:second_sub}
    }
    \caption{An illustration of the block structure for $\cal X \subset\mathbb{R}$ and $k=4$, where there is one (non-zigzag) set of arrivals of size $2k$ between two zigzags and they are indexed by order of arrival. Arrivals with the same color have the same exclusion radius and file size (height). In the left figure, the base of a rectangle indicates the arrival time. In the right figure, for $W_m\geq\widehat{T}_m-T_m$,  an equivalent representation of block $m$ is depicted. At the beginning of the service of the block, all arrivals are already present, hence the block is in effect maximally compressed in time while still preserving the local FCFS mechanism.}
    \label{fig:block}
\end{figure}

Here are a few observations about this definition:
\begin{itemize}
\item By the definition in (\ref{nm}), the size of each block $Y_m,m\geq 1$, is an integer multiple of $2k$. 
\item 
The index of the $m+1$-st block is an integer 
larger than or equal to $n_m+2k$ (it is equal when 
there are two consecutive zigzag events, with their centers separated by $2k$).

\item {A block begins with a “zag” sub-block of size/length $k$, followed by $\beta_m/(2k) -1$ “intermediate’’ sub-blocks (each of length $2k$, none of which being a zigzag event), and ends with a “zig’’ sub-block of length $k$.
All these sub-blocks are mutually independent, with intermediate blocks being identically distributed. There is no intermediate sub-block if $\beta_m/(2k)=1$.}
\item These blocks are mutually independent and identically distributed, except for $Y_0$. The block $Y_0$ differs from the others only in the first sub-block of size $k$, because its elements have the unconditional distribution.
\end{itemize}

The following proposition is
a summary of what has to be retained from the last observations:
\begin{Prop}  
\label{prop1}
The sequence of random elements $\{Y_m\}_{m\ge 1}$ is i.i.d. and independent of $Y_0$.
\end{Prop}

\begin{Rem}
There are several other options for defining zigzag events while preserving the properties
identified above and below. One of them is that with no-doubling of the last event in the zig.
For example, if $k=1$ (i.e., we have customers with an exclusion set equal to the whole space),
then we do not have to double this event and all the properties identified still hold. This is the case when $\{R_i\}$ is i.i.d. with a distribution with unbounded support, e.g., exponential.
\\
The specific structure proposed here is retained for its simplicity and symmetry.
This concept comes from tropical mathematics and is linked to what is now called the tropical matrix rank.
See, e.g., \cite{Mairesse}.
\\
Note also that, in Subsection \ref{ss:maes}, {Assumption A3}, we may add for free a further assumption (e) to assumptions (a)--(d):
\begin{itemize}
    \item{(e)}
${\mathbb P} (B_i\cup B_j \subseteq S) =0$, for all $i\neq j$.
\end{itemize}
Indeed, if (e) fails for some ~$i\neq j$, then we may merge these two sets and consider a
new collection of $(k-1)$ events $B_i$'s instead.
\\
There are other options for defining the blocks.
For example, we may deal with smaller blocks, by defining $\{n_m\}$ inductively as
\begin{align}\label{nm-2}
 n_{m+1} = \min \{n \ge n_m + 2k : \ {\mathbf I}(C_n)=1\}.
 \end{align}
In this case, the number of arrivals in each block does not need to be a multiple of $2k$. However, we prefer the definition in \eqref{nm} because the variant with \eqref{nm-2} leads to a more complicated construction.

\end{Rem}

\subsection{Reduction to a Lindley-Type Recurrence Relation}
\label{ss:lind}

In view of the service discipline and the definition of blocks, customers are
served {\em block by block}, i.e., the service of the customers
of block $m+1$ may start only when the service of the 
closing customer of block $m$ is completed and, at that time, all customers of block $m$ have already left. 
This can be exploited to derive a Lindley-type evolution equation.

For any customer, define its {\em waiting time} as the time that elapses between its
arrival time and the epoch when it starts its service.
Let $D_m$ denote the departure time of the closing customer of block $m$.
Thanks to the block-by-block service property,
the waiting time $W_{m+1}$ of the opening customer of block $m+1$ is equal to $(D_m-T_{m+1})^+$.

We have $D_m=T_m+W_m+\sigma_m$, where
$\sigma_m$ is the time to serve all customers of block $m$ (or equivalently the
time to complete the service time of the closing customer of this block) when its opening customer starts
its service at time $T_m+W_m$. The key observation now follows from the construction
by induction of Section \ref{sec: Description of Dynamics} that
the random variable $\sigma_m$ is a measurable function of $W_m$ and $Y_m$:
$$\sigma_m=\Sigma(W_m,Y_m).$$
It should be clear that $\sigma_m$ and the dynamics of block $m$ depend on the waiting time $W_m$. To see this, consider two subsequent arrivals in this block whose exclusion sets do not
intersect with each other. If $W_m$ is sufficiently large (e.g., $\infty$), then at the beginning of the service (end of waiting), these two arrivals
will be served at the same time. If $W_m$ is small (e.g., 0), on the other hand, then the two
arrivals may be served one by one, each enjoying the maximum transmission rate free
of interference. 

Fig. \ref{fig:block}  illustrates the block structure based on a deterministic set of $\{B_i\}_{i=1}^k$ in $\cal X \subset\mathbb{R}$ for $k=4$. In Fig. \ref{fig:block} (a), the base of a rectangle indicates the arrival time and the height indicates the file size. There is one (non-zigzag) set of arrivals of size $2k$ between two zigzags, indexed by their order of arrival. Arrivals with the same color have the same exclusion radius and file size.  In Fig. \ref{fig:block} (b), for $W_m\geq\widehat{T}_m-T_m$, all arrivals are already present at the beginning of the service of the block, hence the block is maximally compressed while preserving the local FCFS mechanism.

\begin{Prop}
The random variables $W_n$ satisfy the recurrence relation
\begin{equation}
\label{eq: recursion}
W_{m+1} =
\left(W_m+\Sigma(W_m,Y_m)-(T_{m+1}-T_m)\right)^{+}.
\end{equation}
\end{Prop}
\begin{proof}
We have
\begin{align}
W_{m+1} &= (D_m - T_{m+1}  )^{+}\nonumber\\
&=(T_m+W_m+\sigma_m - T_{m+1}  )^{+}\nonumber\\
& = \left(W_m+\sigma_m-(T_{m+1}-T_m)\right)^{+}\nonumber\\
& = \left(W_m+\Sigma(W_m,Y_m)-(T_{m+1}-T_m)\right)^{+}.\nonumber
\end{align}
\end{proof}
Note that $W_m$ is a function of $\{Y_p\}_{p< m}$, whereas, conditionally on
$W_m=W$, $\Sigma(W_m,Y_m)$ and $T_{m+1}-T_m$ are measurable functions of $Y_m$.
Therefore, the dynamics of the sequence $\{W_m\}$ admits the representation
\begin{align*}
W_{m+1} = f(W_m,Y_m). 
\end{align*}
Therefore (see e.g. \cite{Bor}, Chapter 3, Section 11),
\begin{Prop} The sequence
$\{W_m\}$ forms a time-homogeneous Markov chain.
\end{Prop}

Let 
$$\quad \Sigma_m(x) = \Sigma(x,Y_m) \quad \text{and} \quad \widehat \sigma_m=\Sigma(\infty,Y_m).$$
Two observations are in order:
\begin{itemize}
    \item Each of the maps $x\to \Sigma_m(x)$
    from $\mathbb R^+$ to $\mathbb R^+$ can be built using the construction of the last section as a piecewise affine random stochastic process. In particular, the last map is measurable.
    \item The family of stochastic processes $\{\Sigma_m(.)\}_m$ is i.i.d. 
\end{itemize}
The random variable $\widehat \sigma_m$ can be interpreted as follows. 
Assume that all customers of block $Y_m$ are already available at time $D_{m-1}$ of departure of the last customer from the previous block. These customers then form a pile of hailstones.
Start serving all customers in this pile following the induction steps
of Section \ref{sec: Description of Dynamics}, namely in the order prescribed by the precedence
relations between these customers, and at any time, at the individual rate prescribed
by the induction steps.
Then { $D_{m-1} +\widehat \sigma_m \equiv D_m$ }is the epoch when the last element of this pile leaves.

Furthermore, since the random variables $\{Y_m\}$ are independent for $m\ge 0$ and identically distributed for $m\ge 1$,
the same holds for $\widehat{\sigma}_m$: they are independent for $m\ge 0$ and identically distributed for $m\ge 1$. 

\begin{Lem}
\label{lem10}
Under the foregoing assumptions, $\mathbb E \widehat \sigma_m <\infty,$
for any $m$. Moreover, there exists an absolute bound, say $K$, such that 
${\mathbb E} \Sigma_m(x) \le K$, for all $x$ and $m$. In addition, the random variables
$\{\Sigma_m(x), m\ge 1, x\ge 0\}$ are uniformly integrable and, for any $m$, we have  
$\Sigma_m(x) \to \widehat{\sigma}_m$ a.s., as $x\to\infty$. Therefore, for all such $m$,
${\mathbb E} \Sigma_m(x) \to {\mathbb E}\widehat{\sigma}_1$ as well.
\end{Lem}

\begin{proof}
{As shown in Lemma \ref{lem: geometric}, 
the number of customers in block $m$, $\beta_m$,
is a geometrically distributed
random variable with parameter  $p_b$, multiplied by $2k$.}
Further,
since the total service rate (of all customers) in a non-empty system is bounded
below by a positive constant, say $\kappa$ (see Lemma \ref{lemma1}), the total time for service of all customers
in a block cannot exceed the sum of their heights, divided by $\kappa$. Note also that the random variables 
\begin{align*}
    \sum_{i=n_m-k}^{n_m-1} h_i, \ \ m=1,2,\ldots
\end{align*}
are i.i.d.

Then, for any $m=1,2,\ldots$, 
\begin{align*}
  {\mathbb E} \sum_{j=0}^{\beta_m/(2k)-1} \sum_{r=0}^{2k-1} h_{n_{m}+ 2kj+r} 
&=
{\mathbb E} \sum_{j=1}^{\beta_m/(2k)} \sum_{r=-k}^{k-1} h_{n_{m}+ 2kj+r},  
\end{align*}
where on the left-hand side we have the sum of the heights of customers within one block, and on the right-hand side the first 
$k$ terms are replaced by the first 
$k$ terms from the following block. For the right-hand side, given any value of $n_m$, the random variable $\beta_m/2k$ is a stopping time for the random variables $H_j := \sum_{r=-k}^{k-1} h_{n_{m}+ 2kj+r}$.
{{In particular, for any $n$, the event $\{\beta_m\leq n\}$ does not depend on the random variable $H_{n+1}$.}} Therefore, by the Wald identity, for $m=1,2,\ldots$,  
{{
\begin{align*}
{\mathbb E} \sum_{j=0}^{\beta_m/(2k)-1} \sum_{r=0}^{2k-1} h_{n_{m}+ 2kj+r} 
&=
{\mathbb E} \sum_{j=1}^{\beta_m/(2k)} \sum_{r=-k}^{k-1} h_{n_{m}+ 2kj+r}\\
&=
\frac{{\mathbb E} \beta_m}{2k} {\mathbb E} \sum_{r=0}^{2k-1} h_{r}
= {\mathbb E} \beta_2 {\mathbb E} h_1 =: C_1 <\infty.
\end{align*}}}

In the first block, only the distribution of the $k$ random variables
$\{ h_i\}_{i=0}^{k-1}$,
differs from that of the other blocks, in the sense that these random variables have the initial (unconditional) distribution. Then, by similar arguments,
{{\begin{align*}
    {\mathbb E} 
    \sum_{j=0}^{\beta_0/(2k)-1} \sum_{r=0}^{2k-1} h_{ 2kj+r} 
    =: C_2 < \infty.
\end{align*}}}

Furthermore, for any $m$ and $x$, the value of $\Sigma_m(x)$ cannot exceed $D_m-D_{m-1}$
(recall that $D_m$ is the departure time of the last customer in block $m$) and that
\begin{align}\label{DD}
 D_m-D_{m-1} \leq T_m - T_{m-1} + \frac{1}{\kappa}\sum_j h_j =: Z_m  ,
\end{align}
where the sum is taken over all customers $j$ from block $m$.
Therefore, 
{{\begin{align*}
    {\mathbb E} \Sigma_m(x) \leq {\mathbb E} (T_m-T_{m-1}) + C_1/\kappa =
    2k{\mathbb E} t_1/p_b + C_1/\kappa =: K <\infty. 
\end{align*}}
Next, the uniform integrability of all $\{\Sigma_m(x)\}$ follows directly
because they are all stochastically bounded by $Z_m$ and with finite expectations.}

Finally, we have the a.s. convergence of $\Sigma_n(x)$ to $\widehat{\sigma}_n$ because,
in fact, a stronger ``coupling-convergence'' takes place:
\begin{align*}
\Sigma_m(x) = \widehat{\sigma}_m, \ \ \text{ for all} \ \ x\geq \widehat{T}_m - T_{m}.
\end{align*}
Indeed, given $W_m=x$, the first customer of block $m$ starts its service at time $T_m+x$. So, if $x\geq \widehat{T}_m-T_m$, then all customers of block $m$ are already in the system by time $T_m+x$, like in the saturated system.
\end{proof}

\subsection{Main Stability Results}
\label{subsec: Main Stability Results}

We obtain two stability results, one for the Markov chain $\{W_n\}$ 
and one for the continuous-time process
$\{W(t), t\ge 0\}$ (defined later), which admits $\{W_n\}$ as an embedded Markov chain.

Below, we use the following notation: $T:=T_2-T_1$, $\beta:=\beta_1$ (where the latter
is defined in the proof of Lemma \ref{lem10}), and $\widehat{\sigma}:=\widehat {\sigma}_1$.
{{Since $T= t_{n_2}-t_{n_1}$, ${\mathbb E} \beta =2k/p_b$,$\Lambda = 1/\mathbb E t_1$}} and $\beta$ does not depend on $\{t_i\}$, we get
{{\begin{align}
{\mathbb E} T = \frac{2k}{\Lambda p_b}.
\end{align}}}

Our first stability result is as follows:
\begin{Th}
\label{thm1}
Under the foregoing assumptions, 
\begin{enumerate}
\item[(i)] If 
${\mathbb E} T > {\mathbb E} \widehat \sigma$ -- or, equivalently,
{{\begin{align}\label{eq:crit}
\Lambda < \Lambda_c := \frac{2k}{p_b {\mathbb E} \widehat{\sigma}},
\end{align}}}
then there exists
a finite $K>0$ such that the set $[0,K]$ is {\em positive recurrent},
i.e., $\tau_x \equiv \tau_x(K) = \min \{n\ge 1: W_n \le K\} <\infty$
a.s., for any $W_0=x$ and, moreover,
\begin{align*}
\sup_{0\le x \le K} {\mathbb E} \tau_x < \infty.
\end{align*}
In addition, 
the Markov chain $\{W_n\}$ admits a unique  stationary distribution, say $\pi$,
and, for any initial state $x\in {\cal X}$, the distribution of the
Markov chain converges to $\pi$ in the total variation norm.

\item[(ii)] If ${\mathbb E} T < {\mathbb E} \widehat \sigma$ and
if there is a constant $\gamma>0$ such that, for any $x$
\begin{align}\label{instab}
{\mathbb P}(\Sigma(x,Y_1)>T+\gamma)\geq \gamma,
\end{align}
then,  
for some $\varepsilon >0$ and for any initial $x\ge 0$, we have
\begin{align*}
\liminf_{n\to\infty} \frac{W_n}{n} \ge \varepsilon \quad \text{a.s.}
\end{align*}
and, in particular, $W_n \to \infty$ a.s. as $n\to\infty$.
\end{enumerate}
\end{Th}

\begin{remark}
\label{Rem:general-renewal-1}
The statement of Theorem \ref{thm1} continues to hold true in a more general setting, for a renewal arrival process, if the distribution of $t_1$ has unbounded support.
\end{remark}

\begin{remark}\label{Rem12}
    In some special cases, there are closed-form expressions for $p_b$ and ${\mathbb E} \widehat  \sigma$. {For example, for ${\cal X} =[-Q,Q]^2$, $S_i=B(x_i,R_i)$ where $\{R_i\}$ are i.i.d. exponentially distributed random variables with mean  $R$}, it is sufficient to define the zigzag event based on \textit{a single arrival} for which the exclusion set covers $\cal X$, that is, $C_n = \{{\cal X}\subset S_n\}$. In this case, $p_b$ is simply the tail probability of an exponential random variable, and we have:
    {{
    	$$\Lambda_c= \frac {\exp(\sqrt{2}Q/R)}{{\mathbb E} \widehat  \sigma}.$$}}
The tractability of ${\mathbb E} \widehat  \sigma$ depends both on the distribution of $S_i$ and the pathloss function $\ell$.  In general, we can estimate $p_b$ and ${\mathbb E} \widehat  \sigma$ by Monte Carlo simulation. The details are given in Section \ref{sec: Simulations}.
\end{remark}

For any time $t\ge 0$, consider an auxiliary system in which we stop arrivals from time $t$ onward and admit only customers arriving before time $t$. In this system, there is a (random) finite number of arrivals. Denote by $Z(t)$ the time at which the last customer leaves such a system and by $W(t)=Z(t)-t$ the time to empty the system after the interruption of the input process. One can see that $Z(t)$ is finite a.s. (in particular, because the sum of the service rates of customers on the ground is bounded from below) and that $W_n=W(T_n)$, for $n=1,2,\ldots$.

Our second stability result is:
\begin{Th}
\label{thm2}
The following hold:
\begin{itemize}
    \item[(i)] 
    Assume the assumptions of item (i) of the previous theorem hold. The stochastic process $W(t)$ admits a single stationary distribution and, for any initial finite configuration of customers, the distribution of $W(t)$ converges to its stationary version in the total  variation norm.
    \item[(ii)]
    Under the assumptions of item (ii) of the previous theorem, there exists $\delta >0$ such that, for any initial value of $W(0)$,
    \begin{align*}
        \liminf_{n\to\infty} \frac{W(t)}{t} \ge \delta \quad \text{a.s.}
    \end{align*}
\end{itemize}
\end{Th}

The proofs of these two theorems are given in Appendix \ref{app1}.

\begin{remark}
\label{Rem:general-renewal-2}
The statement of 
Theorem \ref{thm2} continues to hold true for a renewal arrival process if the distribution of $t_1$ has a spread-out distribution.
\end{remark}

\begin{remark}\label{Rem13}
There are three stochastic processes of interest.
Two are in discrete-time and one is in continuous-time.
The continuous-time process is studied in Theorem \ref{thm2}.
It roughly describes the evolution of workload in continuous time.
The first discrete-time sequence is
the Markov chain studied in Theorem \ref{thm1}. It is obtained when
sampling this continuous-time process at discrete epochs which are
arrival epochs with an arrival event index $n$ that is a multiple of $2k$ and
such that time $t_n$ is the center of a zigzag event.
There is of course another natural discrete-time stochastic process, which
is that of the sequence of workloads at all arrival epochs.
It should be clear that, under the stability condition, this last discrete-time sequence
admits a unique stationary regime which is reached from all initial conditions.
This last stationary regime is obtained from that of the Markov chain of Theorem \ref{thm1}
using the exchange formula of Palm calculus.
\end{remark}

%% file: Simulations.tex
\section{Numerical Evaluations}
\label{sec: Simulations}

{{In this section, we provide numerical evaluations of the system performance and compare the stability threshold with relevant models. For this comparison, we recall from Subsection \ref{ssap} that $\lambda = \Lambda /m(\cal X)$ and define $
    \lambda_c= {\Lambda_c}/{m(\cal X)}$.}}
\subsection{Estimates of the Stability Threshold}
\label{ss:lc}
In view of the expression obtained in (\ref{eq:crit}),
the most natural way to estimate $\lambda_c$ is to estimate
$p_b$ and $\mathbb E \widehat \sigma_1$ by Monte Carlo simulation.
Both problems can be solved by producing perfect block samples and
using them as input to the induction steps of Section \ref{sec: Description of Dynamics}.
For this purpose, in these blocks, there is no need to keep the arrival epochs.
The simplest way is to first sample an i.i.d. sequence of arrival loci, height, and exclusion sets 
until one sees a first $C$ event and to proceed sampling such data until one sees a second $C$ event.
This provides a sample of a block by restriction to arrivals 
that start with the zag of the first $C$ event and last until the zig of the second one.
This provides a sample of the number of arrivals in a block, {say $\beta$. 
Then we take this block as input to the induction steps of Section \ref{sec: Description of Dynamics}.
This in turn provides a perfect sample of $\widehat \sigma_1$.
By repeating this, one gets an estimate (together with confidence intervals)
of $\mathbb E \beta= 2k/{p_b}$ and
of $\mathbb E \widehat \sigma$.}

\subsection{Examples}
This subsection defines two sets of basic examples, one in the continuous space setting and the other one in the discrete space setting.

\subsubsection{Continuous-Space Setting}
\label{subsubsec: example_continuum}
Consider the PHWG model with a
state space equal to the square window {${\cal X}= [-Q,Q] ^2$ and $m$ the Lebesgue measure on the latter.} The torus version of the model is used to avoid boundary effects. 

We consider exclusion sets $\{S_i\}$ of two types: (1) balls of random radii centered at the point of arrival, where 
the random radii are i.i.d. exponential random variables of mean $R$; (2) balls of fixed radius $R$ centered at the point of arrival. In both cases, the heights $\{h_i\}$ are i.i.d. and exponentially distributed.

We simulate the example defined on the torus version of {${\cal X}= [-Q,Q] ^2$}, $b=1, s = l(0)$. We compare the stability region (captured by the critical threshold) with different fixed or i.i.d., exponentially distributed radii, for a fixed function $l$. 

We first give the analytical results on the two special cases of exclusion sets. For $R\equiv0$, the model is a WSBD process defined in \cite{sankararaman2017spatial}. The stability region is {$\lambda<\lambda_c$}, where the critical threshold is 
\begin{equation}
    \lambda_c = \frac{l(0)}{\ln(2)\mathbb{E} h \int_{\cal X}l(\|x\|)\mathrm{d}  x}. 
    \label{eq: lambda_c^0}
\end{equation}
On the other hand, the dynamic reduces to that of an M/M/1
queue with global FCFS when the exclusion sets are such that there is at most one customer on the ground at any time.
For such an M/M/1 system, the stability region is also known as {$\lambda<\lambda_c$}, and
{{\begin{equation}
    \lambda_c = \frac{c}{\mathbb E h}, 
    \label{eq: lambda_c^W}
\end{equation}}}
where $c$ is the constant denoting the maximum Shannon rate achieved when there is no interference, as defined earlier. 



\begin{figure}
    \centering
\includegraphics[width=0.5\linewidth]{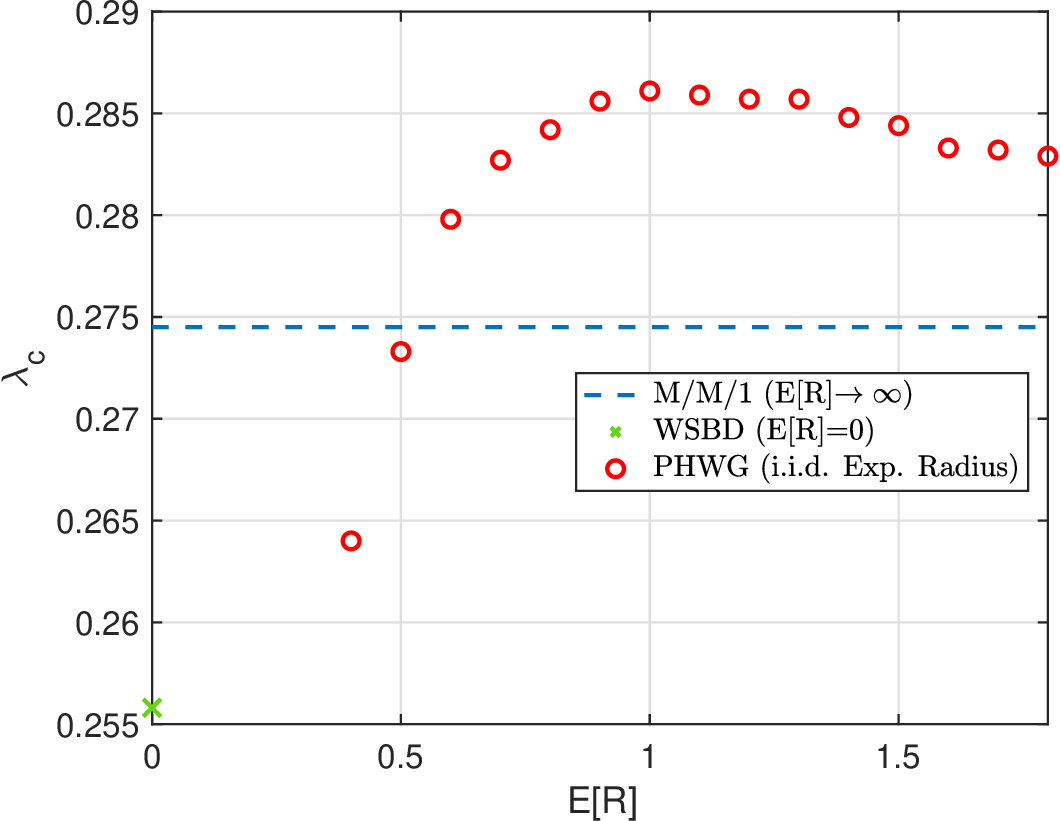}
    \caption{$\cal X$ = $[-2,2]^2$. $\mathbb{E} h = 1$. The exclusion sets have i.i.d. exponential radius, and function $l(r) = \min(1, r^{-4}), w=0.05$.}
    \label{fig:lambda_c_exp_radius}
\end{figure}

\begin{figure}
    \centering
\includegraphics[width=0.5\linewidth]{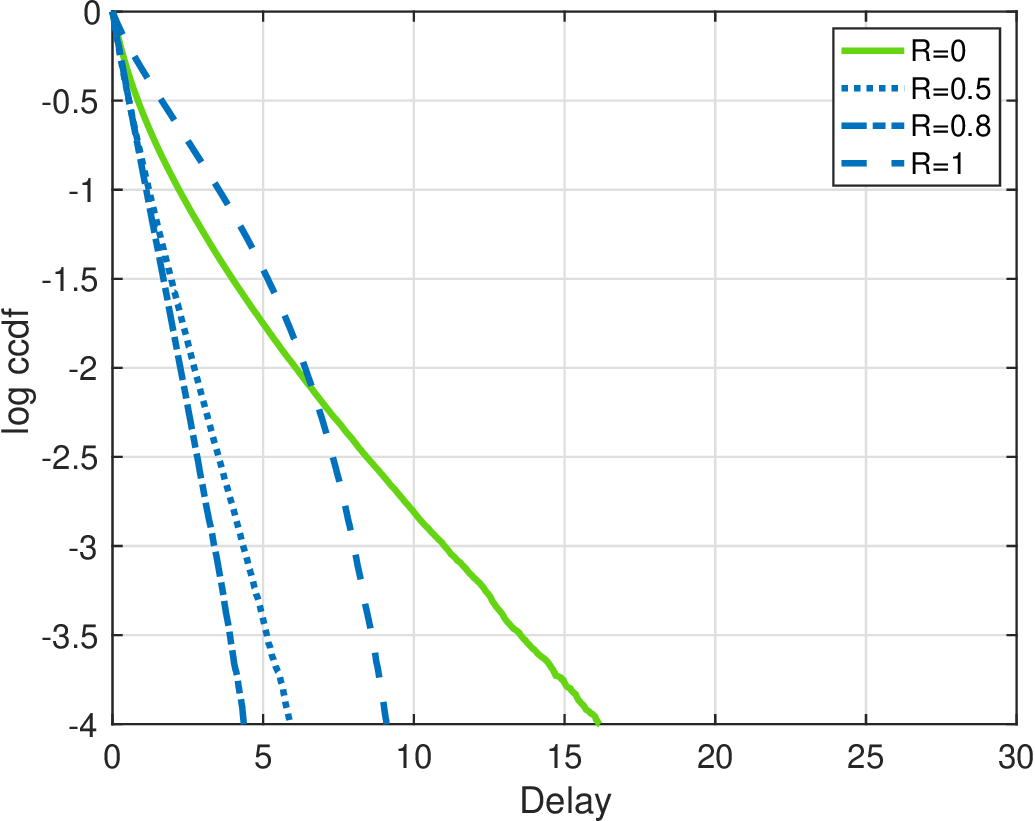}
    \caption{The log ccdf of delay (time elapsed between the arrival and departure of customers) for different constant radii. $\cal X$ = $[-10,10]^2$. $\lambda=0.15$. $\mathbb{E} h = 1$ and $l(r) = \min(1, r^{-4}), w=0.05$.}
    \label{fig:lambda_c_cst_radius}
\end{figure}

In Fig. \ref{fig:lambda_c_exp_radius}, we consider ${\cal X} = [-2,2]^2$, $\mathbb{E} h = 1, w=0.05$, and the function $l(r) = \min(1, r^{-4})$, with i.i.d. and exponential radii for the exclusion sets.  We plot the critical threshold versus the mean radius. From Eq. (\ref{eq: lambda_c^0}) and (\ref{eq: lambda_c^W}), the critical thresholds for the special cases $\mathbb{E} R=0$ and $\mathbb{E} R=\infty$ are 0.2558 and 0.2745 for $R=0$ and M/M/1, respectively. The estimation of the stability threshold is done as described in Subsection \ref{ss:lc} with 1000,000 realizations of blocks (with the exception of the data point at $\mathbb{E} R =0.4$, which is calculated from 5,500 block realizations due to the complexity of the system. The stability threshold, seen as a function of the mean exclusion radius, displays unimodality: $\lambda_c$ has a single maximum which appears at $0<E[R]<\infty$. This leads to what can be referred to as an ``optimal exclusion radius". The results show that, for the given scenario, moderate protection optimizes the system stability as opposed to immediate access, which is too greedy, or global FIFO, which is too conservative.

In Fig. \ref{fig:lambda_c_cst_radius}, we compare the delay distributions of the system with 1000,000 arrivals and arrival intensity $\lambda=0.15$ for different constant radii, $R=0,0.5,0.8,1$. ${\cal X} = [-10,10]^2$, $\mathbb{E} h = 1, w=0.05$, $l(r) = \min(1, r^{-4})$. The case with $R=0$ is equivalent to WSBD. From Eq. (\ref{eq: lambda_c^0}) and (\ref{eq: lambda_c^W}), the critical thresholds for the special cases $ R=0$ and $ R\geq 5.8579$ are 0.23 and 0.011 for $R=0$ and M/M/1, respectively. For $R=0.5$, simulations suggest that $\lambda_c\approx 0.25$.  The delay distribution first improves from $R = 0$ to $R = 0.8$, and then decreases at $R=1$. For constant radii, the delay distributions are consistent with the unimodality of the stability threshold shown in Fig. \ref{fig:lambda_c_exp_radius}. It is also worth noting that the delay distributions for $R=0$ and $R=1$ have different slopes that intersect each other at one point. This is likely because for $R=1$, the service rate is (significantly) higher than $R=0$ on average due to protection while the waiting time is (significantly) longer.






\subsubsection{Discrete-Space Setting}
We now define and analyze an example in the 1D discrete-space setting. Let $N$ be an even natural number $N\geq 4$. This model belongs to the general framework of the paper, with discrete state space and Poisson arrivals.
There are $N$ arrival loci located equidistantly on the circle of length $N$. We number them from $1$ to $N$ clockwise, say.
Customers are represented as {(open)} rectangles $M \times h$ centered at $x$.
Here, $x$ is one of the loci.
Here, $M$ is a fixed integer, with $1 \leq M \leq N$.  
As in the previous discussions,
we assume here that customers are served in a locally FCFS way.
Finally, $h$ is a positive random variable which represents the file size of a customer.

The distance between two customer centers is the minimum of the two distances between them,
clockwise and counterclockwise. Let $d$ be a fixed positive integer.
{The attenuation function at distance $u\in \mathbb N$ is 
$l(u) = 1_{u=0}+ v\cdot 1_{0< u\le d}$, where $v> 0$ is a constant.}
The thermal noise is equal to 1. Hence, the service rate of
a customer at $x$ and in service is equal to ${1}/{(1+v\phi_d(x)})$ 
with $\phi_d(x)$ the number of customers being served at a distance 
less than or equal to $d$, without counting the customer at $x$.

Note that for $M > d$, this model coincides with Poisson Hail since the rate of any customer
being served is always 1. It is known that Poisson Hail is strictly monotonic in the footprint of the hailstones.
More precisely, for $N$ and $d$ fixed, the value of $\lambda_c$ is a decreasing function of $M$, in the range
$M >d$.

Note that if $M>N/2$, then there is no way to serve two or more customers simultaneously, and if $N/2 \geq M > N/3$, then there may be the simultaneous service of at most two customers.

{{
A special case of this model is obtained by taking $d=N/2$ and $N/2 \geq M > N/3$.}
That is, a customer's service ``interferes''  with at least another customer, then it interferes with exactly
one customer, and its service rate is then ${1}/{(v+1)}$.
If $M>N/2$, then there is no way to serve two or more customers simultaneously, and we have a single-server queue
with service rate 1.
If $M=N/2$, then the simultaneous service of two customers may only occur if the distance between the customer
centers is exactly $N/2$, and at most two customers are served simultaneously, with rate ${1}/{(v+1)}.$
}

Consider a particular case $M=d=N/2$.
Consider the saturated system, namely the system with infinitely many customers in the queue in front of it,
which are numbered $1,2,\ldots$. Assume that customer $i$ is centered at $x_i$ and
brings $h_i$ units of work to serve, that $\{x_i\}$ is an i.i.d. sequence with elements uniformly
distributed on the discrete set of $N$ locations, and that $\{h_i\}$ is an i.i.d. sequence of positive random variables.
{{
Let us number customers $0,1,2,\ldots$ Assume that customer $0$ has its center at $x_0$, and let $\nu_1$ be the number of the first customer whose center is located neither at $x_0$ nor at $(x_0+N/2) \mbox{mod}\ N$. Then we define the first block of customers as customers enumerated $0,1, \ldots, \nu_1-1$, and their number, $\nu_1$, has a geometric distribution with parameter ${(N-2)}/{N}$. Further, each customer in the block is centered at $x_0$ with probability $1/2$, independently of anything else. All following blocks are formed similarly. So
}} 
customers are served in blocks of random sizes that are i.i.d. and have a geometric distribution with parameter $p=(N-2)/N$.
Given that there are, say $\nu=n$ customers in a ``typical'' block, a binomial number of them, say $m_n \sim Bin (n,1/2)$, have the total amount of work $H_1:= \sum_1^{m_n} h_i$ and occupy the same $N/2$ servers and $n-m_n$ others occupy the other $N/2$ servers and have the total amount of work $H_2:= \sum_1^{n-m_n}h'_j$ where $\{h'_j\}$ are independent copies of $h_1$.
Then the mean service time of a typical block of customers, say $s$, is
{{
\begin{align*}
s&=
{(1+v)} {\mathbb E} \min (H_1,H_2)
 + {\mathbb E} |H_2-H_1|\\
 & \peq{a} {(1+v)} {\mathbb E} \min (H_1,H_2)
 + {\mathbb E} (H_1+H_2-2\min(H_1,H_2))\\
 &\peq{b} \frac{N}{N-2}{\mathbb E} h_1 +
{(v-1)}{\mathbb E} \min (H_1,H_2),
\end{align*}
where step (a) follows from the fact that $|H_2-H_1| = \max(H_1,H_2)-\min(H_1,H_2) = H_1+H_2-2\min(H_1,H_2)$, and step (b) follows from the linearity of expectation and the mean of the geometric distribution of the number of customers with parameter $(N-2)/N$.

In the particular case where $h_i=1$ a.s., the formula for $s$ becomes slightly simpler:
\begin{align*}
s= \frac{N}{N-2} + (v-1){\mathbb E} \min (m_{\nu}, \nu-m_{\nu}).  
\end{align*}}
Here the critical value $\Lambda_c$ is
\begin{align*}
    \Lambda_c = \frac{N}{(N-2)s.}
\end{align*}
}

%% file: Spatial_Interacting_Queuing_Processes.tex
\section{Spatial Interaction Queuing Processes}
\label{sec: SIBD}
This section defines a new and general class of spatial queuing processes.
Arrivals in these systems are in terms of e.g. a Poisson rain or a renewal
rain, as above. Arrivals are independently marked, with a locus of arrival $x$ in a
given compact $D$ of the Euclidean space, a height $h$ representing the
size of the task, and an exclusion set $B$. The service discipline is locally FCFS,
based on the exclusion sets, with the same definition as above.
The generalization comes from the fact that the interaction function,
which gives the service speed of customers in a given spatial
configuration, is general. More precisely, let $\psi$ denote the point process
of customers currently being served. For all $x\in \psi$,
the service rate of $x$ in this configuration is of the form $s(x,\psi)$,
where $s$ is a measurable function from $D\times {\cal M}_D$ to $\mathbb R^+$.
Here, ${\cal M}_D$ denotes the space of counting measures on $D$.
The Shannon rate function (\ref{eq1}) used above was of this form.
In this special case, the map is monotonic in $\psi$.
In this section, we do not make any assumption of this kind.
We assume that the $s$ function defined above is bounded below by a positive
constant and bounded above by another constant.

It can be seen that the induction of Section \ref{sec: Description of Dynamics} extends
without modifications to this more general setting.
Similarly, blocks can be defined in the same way as in Subsection \ref{ss:lind}.
Using this, one can show that the stability condition of this general class of systems is also of the form 
\begin{align*}
{\mathbb E} T > {\mathbb E}\widehat \sigma
\end{align*}
(with the notation of Section \ref{sec: Stability}), 
with the same interpretation of $\widehat \sigma$ as the time to serve
a block when all elements of the said block are available at time zero
and are served in the FCFS order prescribed by the constraints at the speed
prescribed by the interaction function $s$.

%% file: Discussion_and_Open_Questions.tex
\section{Discussion and Open Questions}
\label{sec:open}
\subsection{On the Complexity of this Class of Systems}
The first set of questions concerns the complexity of the estimation of $\lambda_c$
and of other statistical properties of this class of dynamical systems.
If there are a few cases where one can get closed-form expressions for $\lambda_c$,
even the simulation of a perfect sample of the two random variables
allowing one to evaluate $\lambda_c$ (see Subsection \ref{ss:lc}) may be
a challenging problem. This is well illustrated by the following discrete model: 
the arrival locations are uniformly distributed over the integers ranging from 1 to $N$ on the real line,
and the exclusion radii are equal to $2$ (namely a customer of locus $k$ can only
be served after all customers of locus $k-2,k-1,k,k+1$ and $k+2$ that arrived
before have left - conditions involving integers not in $[1,N]$ are void).
{Assuming that $N$ is odd, a natural collection of sets implementing the
conditions of Subsection \ref{sss:msa} is $B_1$ the ball (of diameter 3) centered
in 2, $B_2$ that centered in 4, etc, up to $B_K$ with $K=\frac {N-1}2$ that
centered in $N-1$. Then the probability of event $C$ is
$\left(\frac 1{N}\right)^{2K}.$
For $N=21$, the mean number of arrivals until event $C$ is 
$21^{20} > 10^{26}$.}

\subsection{Connection to the Saturation Rule}
The results obtained in Section \ref{sec: SIBD} can be seen as a partial extension 
of the saturation rule of \cite{BFsaturation95} to non-monotonic dynamics (in contrast to PHHG for instance). For all
models of spatial interaction queues considered there, the stability region is obtained when
assessing the time to empty the system when blocking arrivals, exactly as in
the saturation rule. 

\subsection{Open Problems}
We now list a few open questions.
\begin{enumerate}
\item The saturation rule was established in the stationary ergodic framework.
In contrast, the results of the present paper require independence assumptions. Can one
relax some of these assumptions and still be able to build the stationary regime?
\item The Poisson Hail model was successfully defined on the infinite Euclidean space.
Can one construct the PHWG model and extend the stability results obtained in the
present paper to non-compact phase spaces? On such spaces,
for most sensible models, there will be no zigzag event of positive probability. Therefore, another mathematical approach will be needed.

\item Can one get bounds or estimates on the stationary system time 
of a customer arriving at a given locus, on the stationary point process of customers
in service, or on the total number of customers waiting?
\item The basic dynamics is fundamentally non-monotonic in that delaying the arrival of a
given customer may lead to an earlier departure for another customer. This is why the
saturation rule cannot be applied. Can one circumvent this
difficulty, e.g., by averaging over permutations of the exclusion sets?
\item 
It would be practically useful to adapt the model described here to a situation
closer to the CSMA/CA setting. This will in particular require the extension of
this model to disciplines different from FCFS, and more precisely to disciplines
based on the carrier sensing and back-off mechanisms used in this context.
Conversely, it would be interesting to devise concrete distributed
algorithms that allow one to implement the model described in the present paper.

\end{enumerate}

\section*{Acknowledgments}

The work of F. Baccelli was supported by the European
Research Council project titled NEMO under grant ERC
788851 and by the French National Agency for
Research project titled France 2030 PEPR réseaux du Futur
under grant ANR-22-PEFT-0010.

%% file: appendix.tex
\section{Appendix}
\label{app1}

\subsection{Discrete time}
{
For $x\geq 0$ and $n=1,2,\ldots$, let 
{
\begin{align}\label{xis}
\xi_n (x) = \Sigma_n(x)-(T_{n+1}-T_n).
\end{align}
}
{
It follows from the earlier observations that the families of random variables 
\begin{align}\label{xiiid}
\{\xi_1(x), x\geq 0\}, \{\xi_2(x), x\geq 0\}, \ldots,
\{\xi_n(x), x\geq 0\}, \ldots \ \ \text{are i.i.d. in} \ \ n 
\end{align}
(where, for each $n$, the trajectory of $\xi_n(x)$ is piecewise linear, with a countable number of jumps). {Then the recursion \eqref{eq: recursion}} may be rewritten in the form:
\begin{align}\label{SD}
X_{n+1} = (X_n + \xi_n(X_n))^+, \ \ n\ge 1,
\end{align}
with $X_n=W_n$, and it is  a time-homogeneous Markov chain.
}}

We reformulate our Theorem 1 in a simpler and more general form, as Proposition \ref{propmain}, and provide its proof. Then Theorem 1 will follow if we let $\xi_n(x)$ be defined as in \eqref{xis}.  

\begin{Prop} \label{propmain}
Let a Markov chain $\{X_n\}$ follow recursion \eqref{SD}, with 
condition \eqref{xiiid} being in force. 
Assume that
the family $\{\xi_1(x)\}_{x\ge 0}$ is uniformly integrable and that
\begin{align}\label{conv-sigma}
\xi_1(x) \rightarrow \xi \equiv \xi(\infty) \ \text{ in distribution, as } \ x\to\infty.
\end{align}  
	\begin{enumerate}
\item[(i)] Let ${\mathbb E} \xi <0$. Then there exists
a finite $K>0$ such that the set $[0,K]$ is {\em positive recurrent},
i.e., $\tau_x \equiv \tau_x(K) = \min \{n\ge 1: X_n \le K\} <\infty$
a.s., for any $X_0=x\ge 0$ and, moreover,
\begin{align}\label{recur}
\sup_{0\le x \le K} {\mathbb E} \tau_x < \infty.
\end{align}
If, in addition, the Markov chain has a positive recurrent atom at point, say, $x_0$ and if ${\mathbb P} (X_1=x_0 \ | \ X_0=x_0)>0$, then there exists a unique stationary distribution, say $\pi$, and 
\begin{align}\label{totvar}
    {\mathbb P} (X_n \in \cdot \ | X_0=x) \to \pi (\cdot), \text{as} \ n\to\infty, 
    \end{align}
    in total variation, for any initial value $X_0=x$.
\item[(ii)] Let ${\mathbb E} \xi_1>0$. If there is a constant $\gamma>0$ such that, for any $x$,
\begin{align}\label{instab1}
{\mathbb P}(\xi_1(x)>\gamma)\geq\gamma, 
\end{align}
and if there exists an integrable random variable, say $\eta >0$ a.s.,
such that, for any $x$, {
\begin{align}\label{extra}
\xi_1(x)\geq -\eta \quad \text{a.s.},
\end{align}
}
then, for any initial $x\ge 0$, we have
\begin{align*}
\liminf_{n\to\infty} \frac{X_n}{n} \ge \varepsilon \quad \text{a.s.}
\end{align*}
and, in particular, 
$X_n \to \infty$ a.s. as $n\to\infty$.
	\end{enumerate}
\end{Prop}
{
\begin{remark}
In the case where ${\mathbb E} \xi = 0$, there may be various scenarios,
including positive recurrence, null-recurrence and transience -- depending on the way 
the distribution of $\xi_1(x)$ converges to the distribution of $\xi$ as $x$ grows.
\end{remark} 
}

Before providing the proof of Proposition \ref{propmain}, we recall the classical  Foster criterion and two of its corollaries.

\begin{Prop} Consider a time-homogeneous Markov chain $\{X_n^{(x)}\}$
on a measurable state space $({\cal X, B_X})$ that starts from the initial state $X_0^{(x)}=x$. 
Let {$L: {\cal X} \to [0,\infty)$} be an unbounded measurable function.
Let
\begin{align*}
\delta_x = L(X_1^{(x)})-L(x).
\end{align*}
Assume that there exists $\varepsilon >0$ and $K>0$ such that
\begin{align*}
\sup_{y: L(y) \le K} {\mathbb E} \delta_y < \infty \quad
\text{and}
\quad
{\mathbb E} \delta_x \le - \varepsilon, \ \ \text{if} \ \ L(x)>K.
\end{align*}
Then the set $B:= \{x: \  L(x)\le K\}$ is positive recurrent.
\end{Prop}

The following two corollaries are straightforward. The proof of the first one may be found, e.g., in Chapter 6 of \cite{FCh}.
\begin{Cor} \label{cor1}
Let $D$ be any subset of the set $B$ of the last proposition.
Denote $\psi_x = \min \{n\ge 1: \ X_n^{(x)} \in D\}$. 
If, in the conditions of the Foster criterion, we have in addition that there exist $n_0\ge 1$ and $p>0$ such that 
\begin{align*}
{\mathbb P} (\psi_x \le n_0) \ge p, \ \ \text{for all} \ \ x\in B,
\end{align*}
then the set $D$ is positive recurrent, too. 
\end{Cor}

\begin{Cor}
\label{cor2}
Under the conditions of Corollary \ref{cor1}, assume in addition that the set
$D$ consists of a single point and that the Markov chain is aperiodic.
Then the Markov chain admits a single stationary distribution, say $\pi$,
and, for any initial state $x\in {\cal X}$, the distribution of the
Markov chain $X_n^{(x)}$ converges to $\pi$ in the total variation norm. 
\end{Cor}

We are now in a position to provide the proof of part (i) of Proposition \ref{propmain}.

\begin{proof} 
Let $\varepsilon := -{\mathbb E} \xi >0$ and
$\Delta (x) = {\mathbb E} (X_1 - X_0 \ | \ X_0=x)$. 
We have
\begin{align*}
\sup_{x\ge 0} \Delta (x) < \infty \quad \text{and} 
\quad
\lim_{x\to\infty}\Delta (x) = -\varepsilon <0.
\end{align*}
Therefore, there exists $K>0$ such that 
\begin{align*}
\sup_{x\in [0,K]} \Delta (x) <\infty \quad \text{and}
\quad
 \sup_{x>K} \Delta (x) \le -\varepsilon/2 <0.
\end{align*}
Then the Foster criterion can be applied.

Furthermore, the existence of a positive recurrent atom and aperiodicity imply the stability result.
\end{proof} 
\begin{remark} In the particular case of Theorem 1, the existence of a positive recurrent atom and aperiodicity  are guaranteed by the unboundedness of the support of the exponential distribution of interarrival times.
\end{remark}

Before giving the proof of statement (ii) of Proposition \ref{propmain},
we recall Theorem 35 in Section 11 of \cite{FCh}
 which is formulated below.
Its proof is a minor modification of the proof of the corresponding result from PhD thesis by \cite{DD}.
and, in turn, the latter result is a slight generalization of the instability theorem in \cite{SFDD}.

Consider a time-homogeneous Markov chain $\{X_n^{(x)}\}$ in a measurable state space $({\cal X, B_X})$ that starts from the initial state $X_0^{(x)}=x$. 
Let {$L: {\cal X} \to [0,\infty)$} be an unbounded measurable function. For $K>0$,
let
\begin{align*}
\theta_x(K) = \min \{n\ge 1 \colon L\left(X_n^{(x)}\right) > K\}.
\end{align*}
Let
\begin{align*}
\delta_x = L(X_1^{(x)})-L(x).
\end{align*}
The result of Theorem 35 from \cite{FCh} may be formulated as follows

\begin{Th}\label{FC} Assume there exist numbers $\varepsilon >0, K>0, M>0$
and an integrable random variable $\eta >0$ a.s. such that
 \begin{enumerate}
 \item
 $\theta_x(K)<\infty$ a.s., for all $x\in {\cal X}$;
 \item
 for all $x\in {\cal X}$ such that $L(x)>K$, 
 \begin{align}\label{upper}
 {\mathbb E} \min (\delta_x, M) \ge \varepsilon
 \end{align}
 and
 \begin{align}\label{lower}
  \delta_x \ge_{st} -\eta.
 \end{align}
 \end{enumerate}
 Then, 
 for any $x\in {\cal X}$,
 \begin{align*}
	 \liminf_{n\to\infty}\frac{L(X_n^{(x)})}{n} \geq\varepsilon \quad \text{a.s.}
 \end{align*}
 and, therefore, 
 \begin{align*}
 {\mathbb P}
 \left(
 \lim_{n\to\infty} L(X_n^{(x)}) =\infty
 \right)
 =1.
 \end{align*}
 \end{Th}

 We are now in a position to prove the second statement of Proposition \ref{propmain}.
 {{
\begin{proof} 
Here $({\cal X, B_X})$ is the positive half-line with the Borel sigma-algebra and $L(x)=x$.
Then condition 1 of Theorem \ref{FC} follows from condition \eqref{instab1},
condition \eqref{upper} from the uniform integrability of $\{\xi_1(x), x\ge 0\}$ and from condition \eqref{conv-sigma},
and condition \eqref{lower} is condition \eqref{extra}. 
\end{proof}
\begin{remark} In the particular case of Theorem 1, condition \eqref{instab1} is condition \eqref{instab} and condition \eqref{extra} follows from the fact that
the random variables $\{T_{n+1}-T_n \}$ are i.i.d. with a finite mean, thus one can take $\eta = T_2-T_1$.
\end{remark} 
}}
 
 \subsection{Continuous time, proof of Theorem 2}

 Proof of (i). This follows from known results for regenerative processes -- see e.g. Theorem 6.1.2 and Proposition 7.3.8 in \cite{Asm} 
 {or Theorem 19.1 in Chapter 4 or \cite{Bor} or Theorem 1 in Chapter 7 of \cite{BorFos}}.

 Proof of (ii). This follows directly from the following observations:
 First,
 \begin{align*}
     W(t) \geq W_{\eta(t)} - Z_{\eta(t)} \quad \text{for any} \quad t\ge 0, 
 \end{align*}
 where $\eta(t) = \min \{ n: T_n \ge t\}$ and  $\{Z_m\}$ are defined in \eqref{DD}. 
 Since $Z_m$ are i.i.d. with a finite mean and since $\eta(t) \to \infty $ a.s., 

 \begin{align*}
     \frac{W(t)}{t} \equiv \frac{W(t)}{\eta(t)} \cdot \frac{\eta(t)}{t} \geq
     \frac{W_{\eta(t)}}{\eta(t)}\cdot \frac{\eta(t)}{t} -
     \frac{\eta(t)}{t} \cdot \frac{\max_{1\leq i \leq \eta(t)} Z_i}{\eta(t)} \quad \text{a.s.}
     \end{align*}
     and, by the Strong Law of Large Numbers, 
     \begin{align*}
         \liminf_{t\to\infty} \frac{W(t)}{t} 
     \geq \varepsilon \cdot \frac{1}{{\mathbb E} (T_2-T_1)} >0 \quad \text{a.s.} 
 \end{align*}
 since
 \begin{align*}
     \frac{\max_{1\leq i \leq n} Z_m }{m} \to 0 \quad \text{a.s., as} \quad m\to\infty.
 \end{align*}